\documentclass[linenumbers,twocolumn]{aa}
\usepackage[utf8]{inputenc}
\usepackage{graphics,graphicx}
\usepackage{epsfig}
\usepackage{url}
\usepackage{wrapfig}
\usepackage{float}
\usepackage{subfigure}
\usepackage{booktabs}
\usepackage{threeparttable}
\usepackage{hyperref}
\usepackage{refcount}
\usepackage{orcidlink}
\usepackage[utf8]{inputenc}

\begin{document}

\titlerunning{Optical observations of HB9 and G159.2+3.3}
\authorrunning{J.-T. Li et al.}

\title{Optical observations of the Galactic SNR HB9 and H~II region G159.2+3.3}

\author{Jiang-Tao Li\inst{1}\orcidlink{0000-0001-6239-3821}\thanks{Corresponding author: pandataotao@gmail.com}
\and
Li-Yuan Lu\inst{1,2}\orcidlink{0000-0002-3286-5346}
\and
Huiyang Mao\inst{1}\orcidlink{0009-0006-7138-2095}
\and
Zi-Qing Xia\inst{1}\orcidlink{0000-0003-4963-7275}
\and
Yang Chen\inst{3,4}\orcidlink{0000-0002-4753-2798}
\and
Ping Zhou\inst{3,4}\orcidlink{0000-0002-5683-822X}
\and
Xin Zhou\inst{1,5}\orcidlink{0000-0003-2418-3350}
}

\institute{Purple Mountain Observatory, Chinese Academy of Sciences, 10 Yuanhua Road, Nanjing 210023, People's Republic of China
\and
Department of Astronomy, Xiamen University, 422 Siming South Road, Xiamen 361005, People's Republic of China
\and
School of Astronomy and Space Science, Nanjing University, Nanjing, Jiangsu 210093, People's Republic of China
\and
Key Laboratory of Modern Astronomy and Astrophysics, Nanjing University, Ministry of Education, Nanjing 210093, People's Republic of China
\and
Key Laboratory of Radio Astronomy, Chinese Academy of Sciences, Nanjing 210023, People's Republic of China
}

\abstract
{We present multi-wavelength observations of the Galactic supernova remnant (SNR) HB9 and the \ion{H}{II} region G159.2+3.3 apparently projected nearby, in order to study their properties and potential physical connections.}
{Confirming the physical connections between SNRs and \ion{H}{II} regions are crucial to understand their origin and interactions with the environment. Optical emission lines are powerful tools to measure the physical, chemical, and dynamical properties of the ionised gas, so could further help us to confirm such physical connections.
}
{We present new optical narrow-band images of G159.2+3.3, as well as long-slit medium resolution optical spectroscopy of both G159.2+3.3 and the SNR HB9 projected nearby. These new data are compared to archival multi-wavelength data to study the properties of the multi-phase ISM in and around these two objects.}
{HB9 is bright in $\gamma$-rays, but the $\gamma$-ray morphology is centrally filled and most of it is not clearly associated with the surrounding molecular clouds. There is a weak apparent connection of HB9 to the IR bright enclosing shell of G159.2+3.3 in $\gamma$-ray. The diffuse Balmer line has almost identical morphology as the radio emission in G159.2+3.3, indicating they two are thermal in origin. Using medium-band high-resolution optical spectra from selected regions of the southeast (SE) shell of HB9 and G159.2+3.3, we found the radial velocity dispersion of HB9 along the slit is significantly higher than the FWHM of the lines. In contrast, these two values are both smaller and comparable to each other in G159.2+3.3. This indicates that the gas in HB9 may have additional global motion triggered by the SNR shock. The [\ion{N}{II}]~$\lambda6583~\text{\AA}$/H$\alpha$ line ratio of both objects can be interpreted with photo-ionisation by hot stars or low velocity shocks, except for the post-shock region in the SE shell of HB9, where the elevated [\ion{N}{II}]/H$\alpha$ line ratio suggests contribution from shock ionisation. The measured electron density from the [\ion{S}{II}] 6716/6730 line ratio is significantly lower in the brighter G159.2+3.3 compared to the SE shell of HB9.}
{Our density estimate suggests that G159.2+3.3, although appearing brighter and more compact, is likely located at a much larger distance than HB9, so the two objects have no physical connections, unless the shock compressed gas in HB9 has a significantly lower filling factor.}


\maketitle

\section{Introduction} \label{sec:Introduction}

Young stars form in giant molecular clouds, ionising the surrounding gas and form \ion{H}{II} regions, then become supernovae (SNe) after death and produce supernova remnants (SNRs). Multi-wavelength observations of \ion{H}{II} regions and SNRs help us understand the star formation and feedback processes. There are more than 8,000 \ion{H}{II} regions discovered in the Milky Way (MW), mostly identified in the infrared (IR) due to the strong extinction in the optical band (e.g., \citealt{Anderson14}). On the other hand, only about 300 Galactic SNRs have been identified, most commonly in the radio band (e.g., \citealt{Green19,Green22}). The birth environment of these SNRs, as well as the physical connection of them to the surrounding \ion{H}{II} regions, are only well studied in a small fraction of them (e.g., \citealt{Chen08,Zhou18,Zhou23}).

\ion{H}{II} regions and SNRs are less commonly observed with large optical telescopes (e.g., \citealt{vandenBergh73,vandenBergh78}), partially due to their distribution close to the Galactic plane so the resulting high extinction, and also because many of these sources have very large angular sizes. Optical observations, however, provide important information on the ionisation, heating, and dynamics of the ionised gas, which are key probes of the star formation and feedback processes. Moreover, many of the \ion{H}{II} regions and SNRs are sufficiently bright and large to be effectively observed with small telescopes using narrow-band imaging or spectroscopy (e.g., \citealt{Parker79,Fesen24}).

HB9 (G160.9+2.6) is a large Galactic SNR (angular size in H$\alpha$ $\gtrsim2^\circ$), with a small (angular size in H$\alpha$ $\lesssim10^\prime$) \ion{H}{II} region G159.2+3.3 apparently projected nearby. The H$\alpha$ image of these two objects was first published by \citet{vandenBergh73}. HB9 appears as thin bright filaments in H$\alpha$, forming a semi-circular shell bright in the southern half, with some more diffuse and fainter filaments forming an enclosed heart-like morphology in the northern side, while G159.2+3.3 appears brighter, more compact, and is located $\sim2^\circ$ north from the southern edge of HB9's shell. \citet{DOdorico77} presented optical spectroscopy observations of the bright western edge of HB9, which reveal the average emission line ratios of H$\alpha$/[\ion{N}{II}]$\approx1.33$, [\ion{S}{II}]$\lambda$6716/6730$\approx1.33$, and H$\alpha$/[\ion{S}{II}]$\approx1.72$. Using the line ratio of the [\ion{S}{II}] doublet, \citet{DOdorico77} obtained an electron density of $n_{\rm e}=690\rm~cm^{-3}$, assuming a temperature of $T\sim10^4\rm~K$.

In the radio band, HB9 appears as multiple bright filaments forming a generally round shape \citep{Leahy07,Gao11}. The average spectral index of the whole SNR is $\alpha\sim0.5$, and $\alpha$ is larger in the interior than on the bright shell, which can be explained by a lower interior magnetic field compared to that near the SNR shock. \ion{H}{I} observations indicate a kinematic distance of $0.8\pm0.4~\rm{kpc}$ for the SNR \citep{Leahy07}, consistent with that derived from optical extinction ($d=0.54~\rm{kpc}$; \citealt{Zhao20}). There is a radio pulsar PSR~B0458+46 located just $\sim23^\prime$ from the centre of HB9, but the latest radio absorption line observation indicates a lower limit of the distance to the pulsar ($2.7\rm~kpc$), which is significantly greater than the distance to the SNR, suggesting it is a background source \citep{Jing23}. The lack of a relation between the SNR and the pulsar also explains their significantly different ages ($\sim6,000~\rm{yr}$ for the SNR, while the spindown age of the pulsar is $\sim1.8\times10^6~\rm{yr}$; \citealt{Leahy07}). Compared to HB9, G159.2+3.3 is much more compact and even brighter in radio, with a morphology resembling the H$\alpha$ image.

HB9 is also detected in X-rays and $\gamma$-rays. With ROSAT observations, \citet{Leahy95} found a centrally brightened soft X-ray morphology possibly enclosed by the H$\alpha$ and radio shells, showing no strong temperature variation across the SNR. The X-ray emission is generally brighter toward the southern H$\alpha$ bright shells. These characteristics are confirmed with deeper Suzaku observations, which further show recombining plasma inside the bright western shell (\citealt{Sezer19}; however, see later analysis showing non-necessity of the recombining plasma; \citealt{Saito20}). Hard X-ray emission above 1.5~keV was detected toward HB9 \citep{Yamauchi93}, but it can be described by a thermal plasma model \citep{Sezer19,Saito20}. Spatially resolved 1-300~GeV $\gamma$-ray emission was also detected with the {\it Fermi} data at a few positions toward HB9, with the strongest features apparently from the SNR interior and associated with the southern shell \citep{Araya14,Sezer19,Oka22}. G159.2+3.3 is not detected in X-ray or $\gamma$-ray in the literature.

\begin{figure*}[hbt]
\centering
\includegraphics[width=1.0\textwidth,trim={0in 0in 0in 0in},clip]{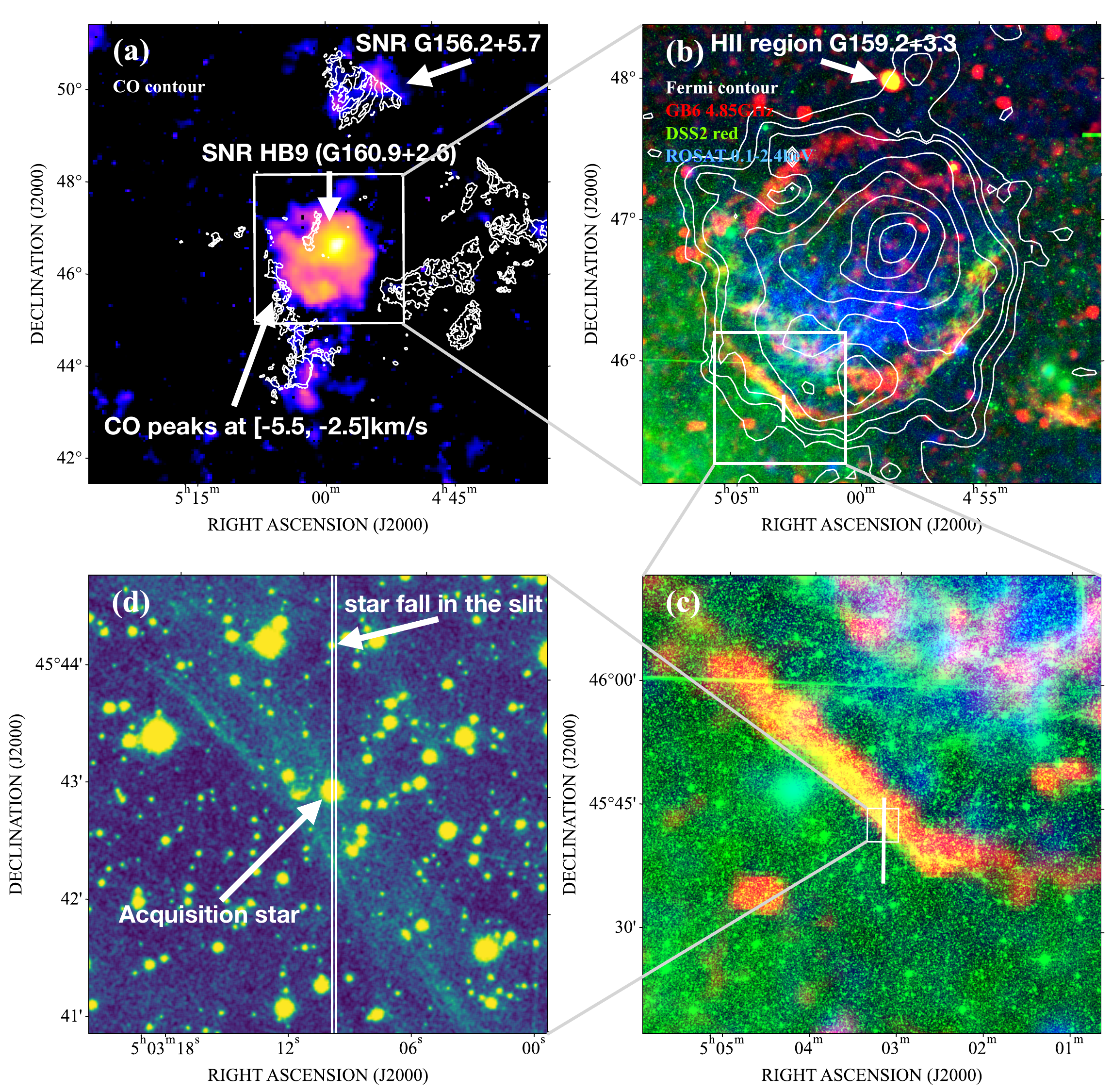}
\caption{Multi-wavelength images of the SNR HB9, the \ion{H}{II} region G159.2+3.3, and the surrounding area. (a) False colour {\it Fermi}-LAT 1-300~GeV test statistic (TS) map centred at the SNR HB9. The white contours are the PMO 13.7m $^{12}$CO (J=1--0) image from the MWISP--CO line survey, integrated in the velocity range of $-13.0 - +6.5\rm~km~s^{-1}$. The $\gamma$-ray and CO enhancement in the upper part of the image is apparently associated with another SNR G156.2+5.7. The $3.25^\circ\times3.25^\circ$ white box in the centre is the FOV of the images shown in panel (b). (b) The red colour is the 4.85~GHz radio continuum image from the Green Bank 6-cm (GB6) survey \citep{Gregory96}. The green colour is the DSS2 red-band image which covers the H$\alpha$ line. The blue colour is the ROSAT/PSPC broad-band (0.1-2.4~keV) image obtained from the ROSAT All Sky Survey (RASS; \citealt{Voges99}). The contours outline the {\it Fermi}-LAT image in (a). The $0.93^\circ\times0.93^\circ$ white box in the lower left is the FOV of (c), with the vertical bar overlaid on the bright southeast filament marks the location of the MDM 1.3m/CCDS slit. (c) A zoom-in of panel~(b), with the slit location plotted as a white bar, which is the same as the white bar in (b). The $3.92^\prime\times3.92^\prime$ black box is the FOV of (d). (d) shows the zoom-in PanSTARRS $r$-band image close to the centre of the slit. The double lines illustrate the width of the slit.}\label{fig:HB9img}
\end{figure*}

\begin{figure*}[hbt]
\centering
\includegraphics[width=1.0\textwidth,trim={0in 0in 0in 0in},clip]{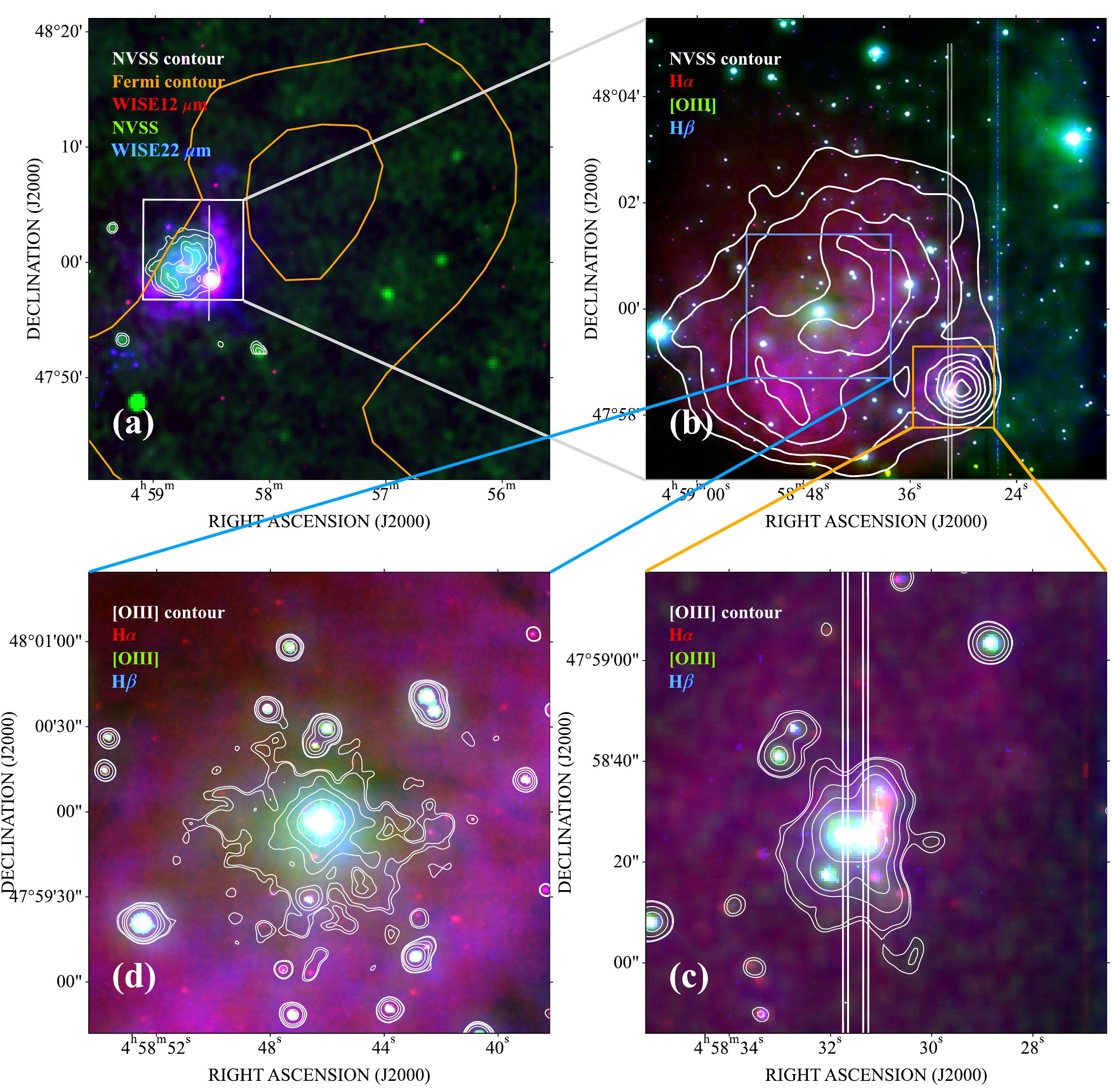}
\caption{Multi-wavelength images of the \ion{H}{II} region G159.2+3.3. (a) NVSS 1.4~GHz (green and white contours) and WISE $12\rm~\mu m$ (red)/$22\rm~\mu m$ (blue) images. The orange contours are the {\it Fermi}-LAT image presented in Fig.~\ref{fig:HB9img}a. The $8.7^\prime\times8.7^\prime$ white box is the FOV of our MDM 1.3m optical images presented in (b), while the vertical bar marks the approximate location of the CCDS slit. (b) Smoothed MDM 1.3m narrow-band images: H$\alpha$ (red), [\ion{O}{III}] $\lambda 5007~\text{\AA}$ (green), H$\beta$ (blue). The contours show the same NVSS image as in (a). The two vertical lines are the locations of the slits used in the two CCDS spectroscopy observations, falling on two close (separated by $\approx3.5^{\prime\prime}$) bright stars, respectively. The smaller orange ($1.52^\prime\times1.52^\prime$) and bigger blue ($2.71^\prime\times2.71^\prime$) boxes are the FOVs of panels (c) and (d), which show two areas with possibly extended [\ion{O}{III}] emission (highlighted with white contours in these two panels).}\label{fig:G159img}
\end{figure*}

We herein present new optical spectroscopic observations of HB9 and the neighboring \ion{H}{II} region G159.2+3.3, as well as new optical narrow-band images of the latter (Figs.~\ref{fig:HB9img}, \ref{fig:G159img}). The major goal is to confirm if there is a physical connection between HB9 and G159.2+3.3 or any other surrounding \ion{H}{II} regions associated with the latter. The paper is organized as follows: in Sect.~\ref{sec:Data}, we introduce our optical observations and data reductions. We further compare our new optical images to the multi-wavelength images, and discuss the scientific implications in Sect.~\ref{sec:Discussion}. The main results and conclusions are summarised in Sect.~\ref{sec:Summary}. Errors are quoted at $1~\sigma$ confidence level in the remaining part of the paper, unless specifically noted.

\section{New optical observations and data reduction} \label{sec:Data}

\subsection{Narrow-band images} \label{subsec:NarrowBandImages}

We took narrow-band images of G159.2+3.3 with the McGraw-Hill 1.3m telescope at the Michigan-Dartmouth-MIT (MDM) observatory (Fig.~\ref{fig:G159img}). We used the $1024\times1024$ Templeton CCD mounted on the F7.5 focus, resulting in a FOV of $8.49^\prime$ and a pixel size of $0.5^{\prime\prime}$. The FOV almost covers the entire \ion{H}{II} region bright in H$\alpha$. We used the OSU656nb3, OSU486.1/3, and OSU5007 filters in the 2.5-inch OSU interference filter set covering the H$\alpha$, H$\beta$, and [\ion{O}{III}]~$\lambda 5007\text{\AA}$ lines. The full width at 50\% peak transmission of these three filters is 40, 27, and $29~\text{\AA}$, respectively, with detailed information, including the transmission curve, available on the MDM website\footnote[1]{\url{http://mdm.kpno.noirlab.edu/inst_osu_filters.html}}. The data used in this paper were taken before the moonrise, with exposures of $5\times600~\rm{s}$ (on 2024.03.21), $5\times600~\rm{s}$ (on 2024.03.19), and $3\times600~\rm{s}$ (on 2024.03.18) for H$\alpha$, H$\beta$, and [\ion{O}{III}], respectively, depending on the weather and the calibration observations.

We used the \texttt{Astropy} affiliated package \texttt{ccdproc} (v1.3.0.post1; \citealt{Craig17}) to reduce the MDM optical images. We used 40 columns of the overscan area to subtract the bias with the tool \texttt{subtract\_overscan}. We then trimmed the overscan region using the tool \texttt{trim\_image}. We used the instrument flats taken with a built-in continuous lamp to correct for vignetting. The instrument flats for different filters were processed using the same bias-subtraction and overscan-trimming procedure, then renormalised and stacked using the tool \texttt{combine}. Each science exposure was flat corrected using the tool \texttt{flat\_correct}, realigned, and combined together by calculating the median using \texttt{combine} to reject cosmic rays (CRs) and hot pixels. Since different science exposures taken on the same night are often quite close to each other on the sky, the prominent bright column of hot pixels is often not removed. We then combined the reduced images in different bands to produce the colour images in Fig.~\ref{fig:G159img}b-d.

\begin{figure*}[hbt]
\centering
\includegraphics[width=1.0\textwidth,trim={0.2in 0in 0.1in 0.in},clip]{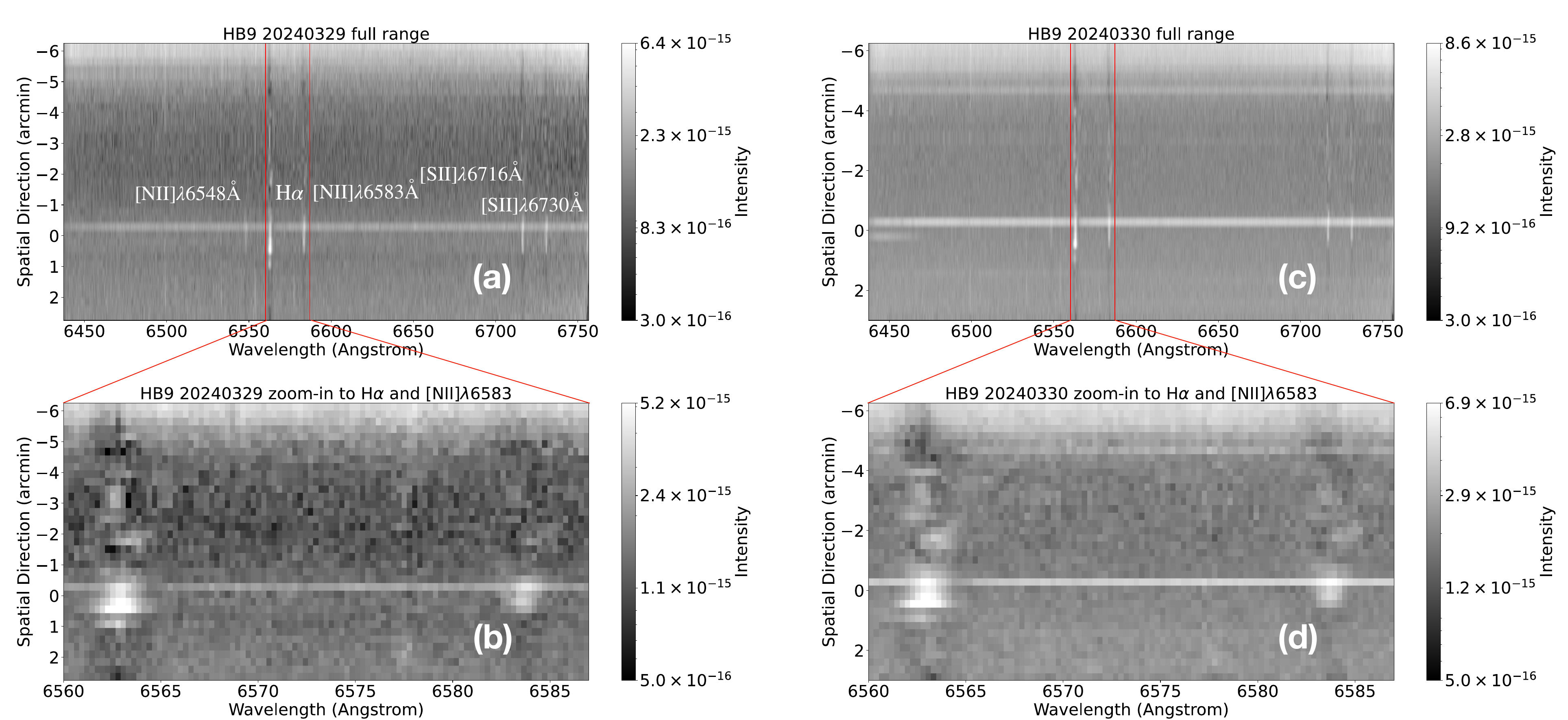}
\caption{The 2D spectra of the SE shell of HB9 extracted from the slit shown in Fig.~\ref{fig:HB9img}. The left and right panels show the spectra taken on 2024.03.29 and 2024.03.30, respectively. The $y$-axis is the distance in arcmin from the bright star used for slit acquisition (appear as a bright horizontal line), while the $x$-axis is the wavelength in \text{\AA}. The colour bar shows the intensity in units of $\rm ergs~s^{-1}~cm^{-2}~\text{\AA}^{-1}$. Prominent emission lines are denoted. The red vertical lines in the upper panels show the wavelength range of the H$\alpha$ and [\ion{N}{II}]~$\lambda 6583~\text{\AA}$ lines, which is further zoomed in in the lower panels. }\label{fig:HB92DSpec}
\end{figure*}

\begin{figure*}[hbt]
\centering
\includegraphics[width=1.0\textwidth,trim={0.2in 0in 0.1in 0.in},clip]{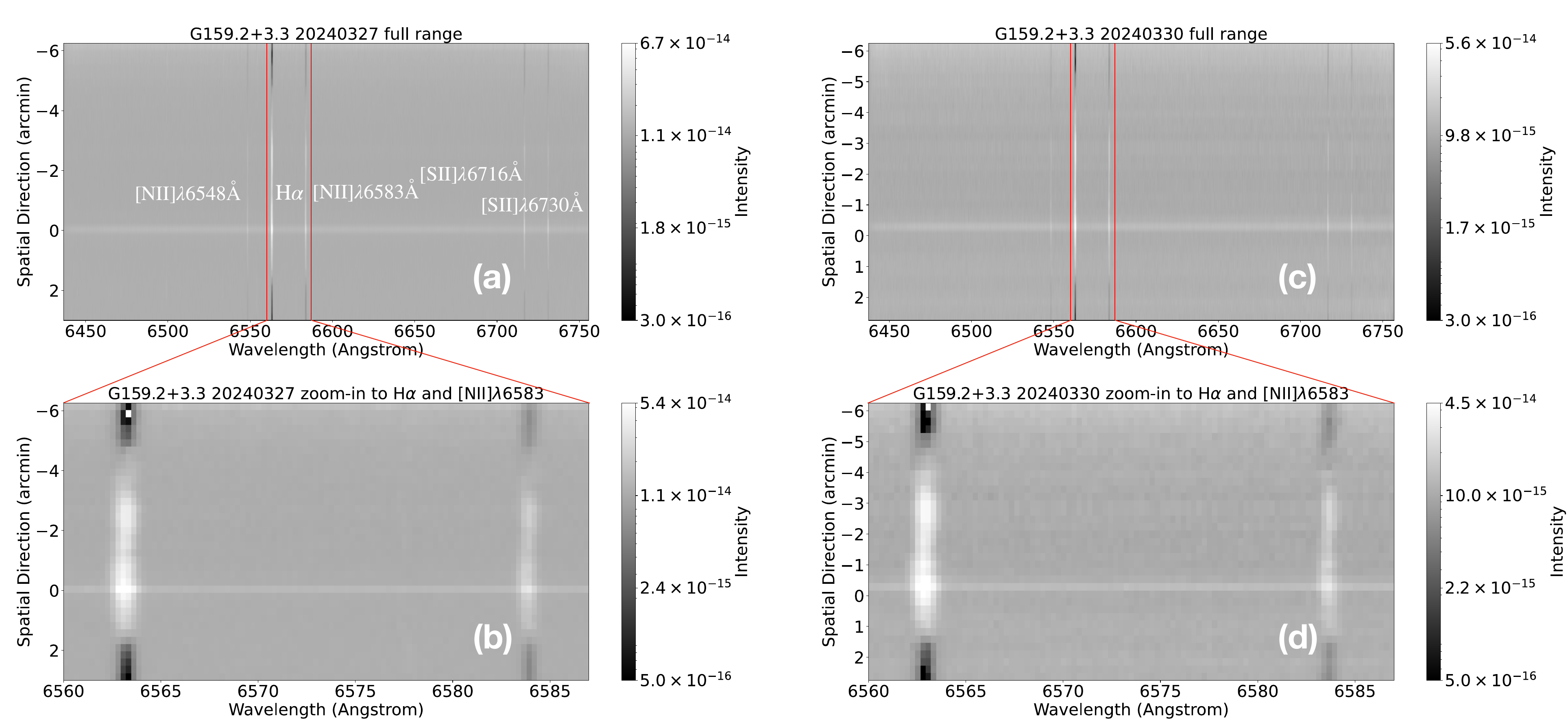}
\caption{Similar as Fig.~\ref{fig:HB92DSpec}, but for G159.2+3.3 extracted from the slits shown in Fig.~\ref{fig:G159img}. The slits used in the two nights (2024.03.27 and 2024.03.30) fall on two neighbouring stars $\approx3.5^{\prime\prime}$ from each other. We do not see a significant difference between the spectra taken in these two nights, even the spectra of 2024.03.30 were taken during the evening twilight with a shorter exposure (Sect.~\ref{subsec:CCDSData}).}\label{fig:G1592DSpec}
\end{figure*}

\subsection{Medium-band spectroscopy} \label{subsec:CCDSData}

We took medium-band optical spectra of a few selected regions from G159.2+3.3 and HB9 with the Boller \& Chivens CCD Spectrograph (CCDS) mounted on the MDM 1.3m telescope. CCDS is a conventional optical grating spectrograph operating in $\sim3,200-9,500~\text{\AA}$. There are six diffraction gratings available on CCDS. We chose the highest resolution 1800~grooves/mm grating, which has a wavelength coverage of $\sim330~\text{\AA}$ on the $1,200\times 800$ CCD, a dispersion of $0.275\rm~\text{\AA}~pixel^{-1}$ ($0.75^{\prime\prime}\rm~pixel^{-1}$ in the spatial direction), and a resolution of $\sim0.3~\text{\AA}$ at $\sim4,700~\text{\AA}$ with a fixed $1^{\prime\prime}$ slit on the 1.3m telescope. We chose a centre wavelength of $\lambda_{\rm cen}\approx 6603~\text{\AA}$, so the spectra cover approximately $6437-6756~\text{\AA}$. This band includes a few important emission lines, such as H$\alpha$~$\lambda6563~\text{\AA}$, [\ion{N}{II}]~$\lambda\lambda6548,6583~\text{\AA}$, and [\ion{S}{II}]~$\lambda\lambda6716,6730~\text{\AA}$.

For HB9, we placed the approximately $10^\prime$-long slit on the southeast (SE) outer shell, on a bright filament probably tracing the SNR's forward shock (Fig.~\ref{fig:HB9img}). This area was close to a $\gamma$-ray enhancement revealed by an earlier {\it Fermi} Large Area Telescope ({\it Fermi}-LAT) observation \citep{Sezer19}, but with the latest {\it Fermi} data, it was identified as a point source and removed (\citealt{Oka22}; Fig.~\ref{fig:HB9img}a). We took two $3\times1200\rm~s$ exposures in two nights (2024.03.29 and 03.30), both with the slit located on a bright star for an accurate target acquisition. The star is $\sim1^\prime$ north of the H$\alpha$ filament (Fig.~\ref{fig:HB9img}d).

For G159.2+3.3, we placed the slit on an IR-bright star ($\rm{RA,DEC=04:58:30.3,+47:58:33}$), in order to better locate the slit and the spectra extracted along it (Fig.~\ref{fig:G159img}). IR emission from this star was detected with the Midcourse Space Experiment (MSX; \citealt{Deharveng05}). The slit length is slightly longer than the FOV of our narrow-band images taken with the MDM 1.3m and Templeton, and covers the eastern edge of the \ion{H}{II} region, including the optical and radio faint, but IR bright outer shell. This IR bright star is resolved into two comparably bright point sources with a separation of $\approx3.5^{\prime\prime}$ on the optical image, with some clearly extended [\ion{O}{III}] emission detected surrounding them (Fig.~\ref{fig:G159img}c). We then placed the slit on these two point sources in different observations. One of the two observations ($3\times600\rm~s$ on 2024.03.30; the other has $3\times1200\rm~s$ on 2024.03.27) was taken during the last half an hour of the evening twilight, so the sky background may be slightly brighter (not quite significant, as shown in Fig.~\ref{fig:G1592DSpec}).  

We reduced the CCDS long-slit spectroscopy data with \texttt{IRAF v2.17} (the Image Reduction and Analysis Facility) \citep{Tody86,Tody93}. We first subtracted the bias with the overscan in the unilluminated columns of 1204-1231 with the \texttt{IRAF} task \texttt{linebias}. We then created the instrument flat with the built-in flat lamp. The CRs were initially cleaned with the \texttt{IRAF} task \texttt{cosmicrays}, and further removed by calculating the median value of three exposures. We chose the MDM 1.3m built-in Xenon arc lamp for wavelength calibration, as it has a few bright lines in the medium wavelength range of interest. We identified these lines using the \texttt{IRAF} task \texttt{identify} and \texttt{reidentify}, then calculated the transformation function using \texttt{fitcoords} and applied it to the raw spectral image using \texttt{transform}. We performed initial background subtraction using the \texttt{IRAF} task \texttt{background}. We next aligned different spectral images using the bright stars falling in the slit, then combined them together. We also made variance images as the measurement error. We applied extinction corrections for different airmasses using the \texttt{IRAF} task \texttt{extinction}. When performing aperture extraction, we first extracted a spectrum only from one bright star using \texttt{apall}. This spectrum was used for a careful background fitting and subtraction, as well as tracing the curvature of the 2D spectrum across the dispersion direction. We then extracted a few spectra along the spatial direction, each with an extent of 20 pixels ($15^{\prime\prime}$). Since most of these spectra were extracted from extended sources with no significant structures in the spatial direction, we adopted the same background and aperture trace models as for the bright star, which was located close to the centre of the slit. The extracted source spectra were then flux calibrated using the spectra of the standard star Feige~34, taken with exactly the same setting. We finally applied the heliocentric velocity correction by calculating it using the \texttt{IRAF} task \texttt{rvcorrect}, then used \texttt{dopcor} to actually shift the spectra. The reduced 2D spectra of HB9 and G159.2+3.3 are presented in Figs.~\ref{fig:HB92DSpec} and \ref{fig:G1592DSpec}, respectively.

We fitted the spectra extracted from individual regions using the curve fitting code \texttt{LMFIT} \citep{Newville24}. We adopted a linear continuum plus five Gaussian models representing H$\alpha$ $\lambda6562.819\text{\AA}$, [\ion{N}{II}] $\lambda\lambda6548.05,6583.45\text{\AA}$, and [\ion{S}{II}] $\lambda\lambda 6716.440,6730.816\text{\AA}$, assuming that they have the same centroid velocity. The full width at half maximum (FWHM) of lines from different elements (H, N, and S) are not linked and could be slightly different. The fitting results, as well as some example spectra, are presented in the Appendix (Tables~\ref{table:HB9specfit} and \ref{table:G159specfit}; Fig.~\ref{fig:ExampleSpec}).

\subsection{Archival multi-wavelength data used for comparison} \label{subsec:MultiwavData}

In addition to the optical images described in Sect.~\ref{subsec:NarrowBandImages}, we also plot a few multi-wavelength images in Figs.~\ref{fig:HB9img} and \ref{fig:G159img} for comparison. We constructed the 1-300~GeV $\gamma$-ray image using more than 15 years (from August 04, 2008 to March 28, 2024) of the {\it Fermi}-LAT \citep{Ajello21} P8R3 data. We adopted the Galactic diffuse emission template, the isotropic diffuse spectral model and the Fourth {\it Fermi}-LAT source catalogue (4FGL; \citealt{Abdollahi20}) to calculate the test statistic (TS) map with a pixel size of $0.05^\circ$ in a $10^\circ \times 10^\circ$ area around HB9. 

We adapted the $^{12}$CO (J=1--0) data from the original one published in \citet{Zhou23}, which were obtained from the Milky Way Image Scroll Painting (MWISP)--CO line survey project \footnote{\url{http://english.dlh.pmo.cas,cn/ic/}} using the Purple Mountain Observatory Delingha 13.7m millimetre wavelength telescope. The half-power beamwidth was $\sim51^{\prime\prime}$, and the data were meshed with a grid spacing of $30^{\prime\prime}$. The typical rms noise level of the integrated intensity in the $1.5\rm~km~s^{-1}$ velocity interval is $\sim0.25\rm~K~km~s^{-1}$. We integrated in a broad velocity range of $[-13.0,+6.5]\rm~km~s^{-1}$ and present the full range $^{12}$CO (J=1--0) image as contours in Fig.~\ref{fig:HB9img}a. 

We also obtained the ROSAT X-ray (0.1-2.4~keV) image, the Green Bank 6-cm (GB6) 4.85~GHz image, and the WISE infrared (IR; 12, $22\rm~\mu m$) images from the SkyView Virtual Observatory \citep{McGlynn98} and plotted them for comparison in Figs.~\ref{fig:HB9img} and \ref{fig:G159img}.

\section{Discussion} \label{sec:Discussion}

\subsection{Comparison of multi-wavelength images} \label{subsec:MultiwavImages}

We compare the multi-wavelength morphology of HB9 and the surrounding area in Fig.~\ref{fig:HB9img}a. This SNR is bright in $\gamma$-ray, but the $\gamma$-ray morphology is not clearly associated with the surrounding molecular clouds (Fig.~\ref{fig:HB9img}a; \citealt{Zhou23}). Instead, some other weaker $\gamma$-ray features in the surrounding area, such as the SNR G156.2+5.7 to the north (e.g., \citealt{Gerardy07}), and a chain of $\gamma$-ray knots to the south, are apparently associated with some molecular gas features. The only molecular cloud apparently associated with the SNR HB9's $\gamma$-ray emission is toward the southeast (SE) shell, where the filament-like molecular cloud appears in a velocity range of $[-5.5,-2.5]\rm~km~s^{-1}$.

As shown in Fig.~\ref{fig:HB9img}b, the SNR HB9 is clearly detected in radio continuum at 4.85~GHz, appearing as an almost complete round shell with a typical diameter of $\sim2^\circ$. The southern half of the shell has generally coherent optical counterparts, while the northern half does not. The SE part of the shell has two layers, both have specifically coherent optical/radio structures. The X-ray emission is relatively hard with little contribution at $\lesssim0.4\rm~keV$, indicating strong extinction toward the SNR ($N_{\rm H}\sim4\times10^{21}\rm~cm^{-2}$; \citealt{Sezer19}). The bulk of the X-ray emission is enclosed on and within the inner SE shell, while there is no significant X-ray emissions along the outer shell. Spectral analysis with the Suzaku data in \citet{Sezer19} indicates that the western part of the X-ray emission is well described by a model having collisional ionisation equilibrium (CIE) and recombining plasma components, while the eastern part is best reproduced by a CIE plus a non-equilibrium ionisation (NEI) model. The $\gamma$-ray morphology revealed by the latest {\it Fermi}-LAT data in this paper and \citet{Oka22} is different from that constructed with the $\sim10\rm~yr$ {\it Fermi}-LAT data in \citet{Sezer19}. The $\gamma$-ray brightest features are still the SNR interior approximately at the centre with no bright multi-wavelength counterparts, as well as the SE inner shell probably coherent with the inner optical and radio shells. However, different from the images presented in \citet{Sezer19}, the SNR interior appears significantly brighter, and the source outside the SE outer shell has now been identified as a point source (4FGL~J0506.6+4545) and removed. Another newly revealed interesting $\gamma$-ray feature is the weak enhancement slightly offset to the west of the \ion{H}{II} region G159.2+3.3 (Figs.~\ref{fig:HB9img}b, \ref{fig:G159img}a). This feature, if real, may be physically connected to the dusty shell and the dense molecular cloud enclosing the \ion{H}{II} region (see below). It may be a signature of the physical connection between these two objects, which however, needs to be further examined. 

In Fig.~\ref{fig:G159img}, we present multi-wavelength images of the \ion{H}{II} region G159.2+3.3 from both our new MDM narrow-band observations and the archival data. It is clear that the IR emission forms a shell enclosing the interior bright in both Balmer lines (H$\alpha$, H$\beta$) and the NVSS 1.4~GHz radio emission. Diffuse [\ion{O}{III}]~$\lambda5007~\text{\AA}$ line is almost non-detectable within the \ion{H}{II} region, except for probably around two bright stars used for target acquisition and a point source at the centre (Fig.~\ref{fig:G159img}c,d). This multi-wavelength morphology is typical for \ion{H}{II} regions (e.g., \citealt{Anderson11,Anderson14}), which strongly indicates that both the Balmer lines and radio emission are thermal; the latter is produced in either radio recombination line or the thermal Bremsstrahlung radiation (e.g., \citealt{Terzian65,Anderson11,Makai17}). On the other hand, forbidden lines and non-thermal radio emissions often elevated in shock, are quite weak. The IR emission from the shell should be mainly produced by the dust heated by the UV photons from young stars, with the $12\rm~\mu m$ emission dominated by polycyclic aromatic hydrocarbon (PAH) molecules, while the $22\rm~\mu m$ emission dominated by small dust grains (e.g., \citealt{Anderson14}). We do not see any significant X-ray emission from G159.2+3.3, consistent with the above scenario that the emission of the \ion{H}{II} region is dominated by the radiation-heated thermal emission with little contributions from shock heating. 

\begin{figure*}[hbt]
\centering
\includegraphics[width=1.0\textwidth,trim={0in 0in 0.3in 0.3in},clip]{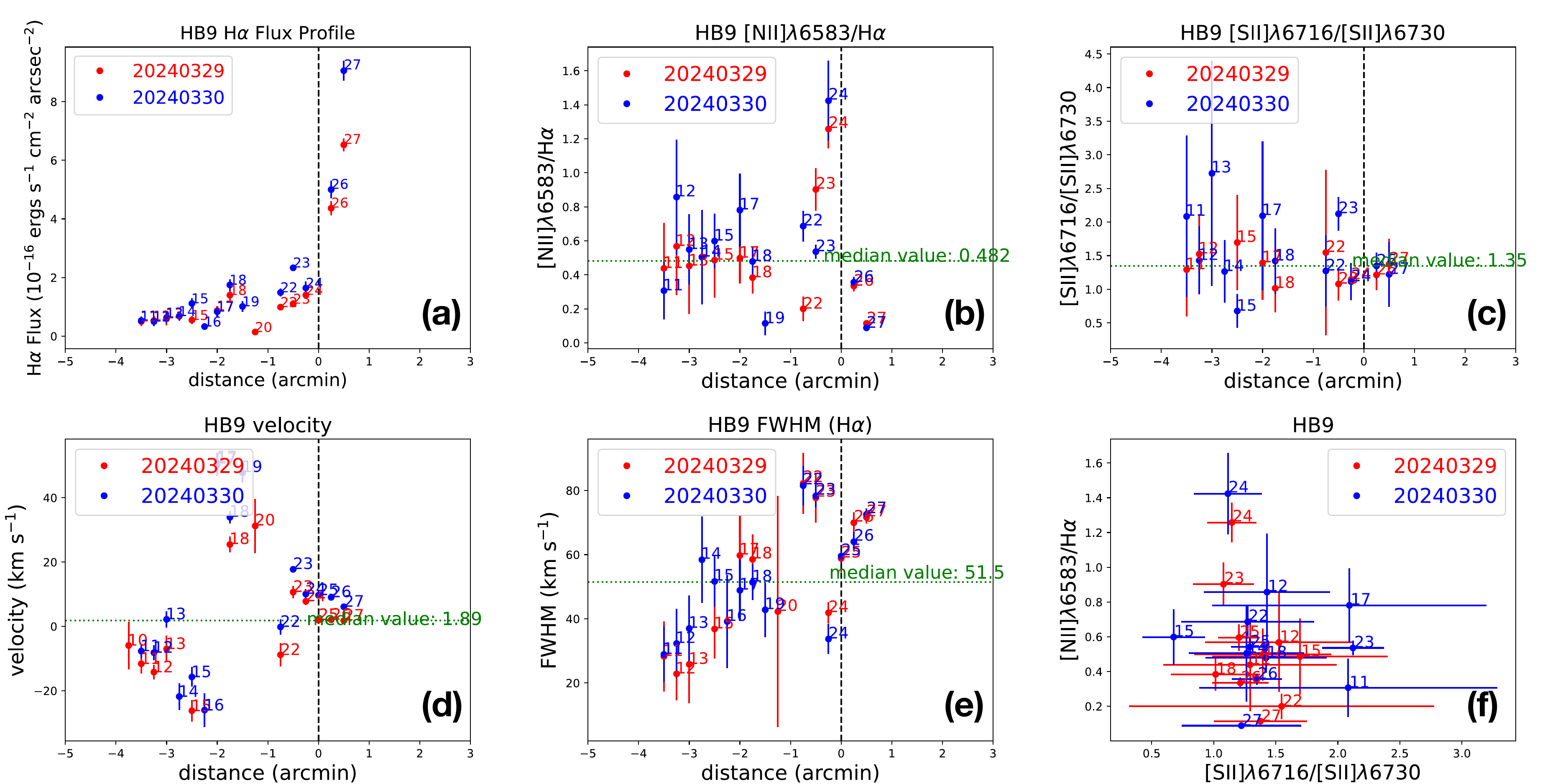}
\caption{Spatial distribution of parameters along the slit of HB9. Panels~(a-e) are the H$\alpha$ flux, the [\ion{N}{II}]$\lambda 6583\text{\AA}$/H$\alpha$ flux ratio, the [\ion{S}{II}]$\lambda 6716/6730\text{\AA}$ flux ratio, the centroid velocity and FWHM of the H$\alpha$ line. Panel~(f) is the [\ion{S}{II}]$\lambda 6716/6730\text{\AA}$-[\ion{N}{II}]$\lambda 6583\text{\AA}$/H$\alpha$ line ratio diagram. The blue and red colours denote the data taken in the two nights, respectively. The distance in arcmin in panels~(a-e) is measured from the bright target acquisition star, with north as negative and south as positive (Figs.~\ref{fig:HB9img}, \ref{fig:HB92DSpec}). In some panels, we also calculate the median value of the parameters (with significant outliers rejected) and denote them with a green dotted line.}\label{fig:HB9profiles}
\end{figure*}

\begin{figure*}[hbt]
\centering
\includegraphics[width=1.0\textwidth,trim={0in 0in 0.3in 0.3in},clip]{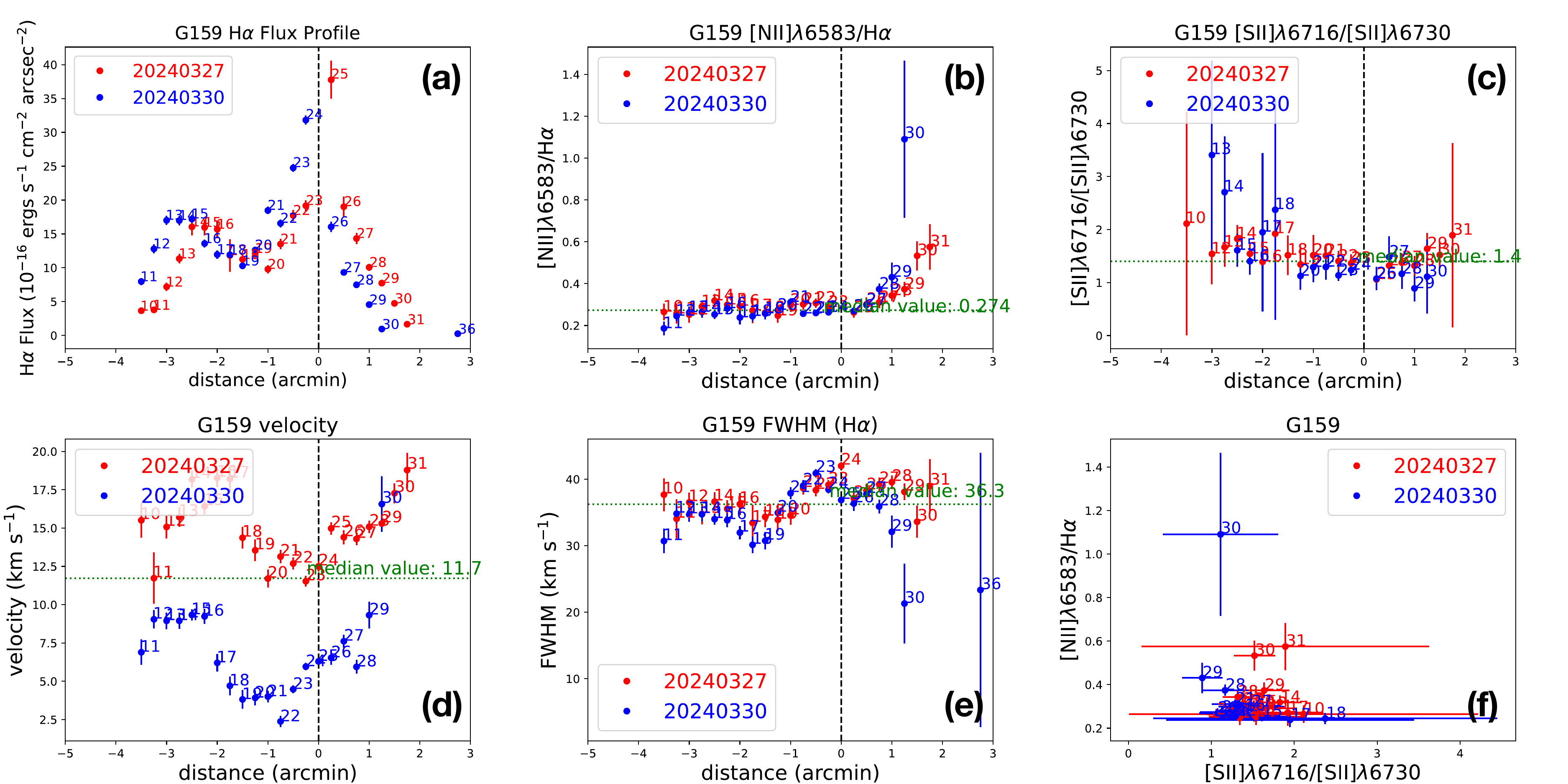}
\caption{Similar as Fig.~\ref{fig:HB9profiles} but for G159.2+3.3. Note that the slits used to extract the spectra in the two nights are slightly offset, located on two close but different acquisition stars (Fig.~\ref{fig:G159profiles}).}\label{fig:G159profiles}
\end{figure*}

\subsection{Radial velocity and gas dynamics} \label{subsec:VelocityDistance}

Measuring the radial velocity of a gas cloud provides a reliable method to estimate its distance, based on the MW kinematic model. \citet{Leahy07} found an \ion{H}{I} 21-cm line emission enhancement spatially associated with only the boundary of HB9 in a local standard-of-rest (LSR) velocity range of $V_{\rm LSR}=-3$ to $-9\rm~km~s^{-1}$, suggesting that the SNR is at a distance of $d=0.8\pm0.4\rm~kpc$. In the MWISP CO survey, \citet{Zhou23} identified only one broad CO line component toward HB9 at $V_{\rm LSR}=-21.6\rm~km~s^{-1}$, which however, shows no good spatial correlation with the remnant. Instead, they found better spatial correlation of the molecular gas with HB9 at $V_{\rm LSR}\sim-2.5\rm~km~s^{-1}$. Furthermore, we also see some CO filaments in the velocity range of $[-5.5,-2.5]\rm~km~s^{-1}$ outside the SE shell of HB9, as discussed in Sect.~\ref{subsec:MultiwavImages} and shown in Fig.~\ref{fig:HB9img}a. The $V_{\rm LSR}=-21.6\rm~km~s^{-1}$ component results in a distance of $d=2.1\pm0.4\rm~kpc$, inconsistent with the estimate from optical extinction \citep{Zhao20}. On the other hand, the $V_{\rm LSR}\sim-2.5\rm~km~s^{-1}$ component indicates a distance of $d=0.5\pm0.2\rm~kpc$, consistent with other estimates \citep{Leahy07,Zhao20}, so could be associated with the SNR.

In optical band, \citet{Lozinskaya81} found a mean radial velocity of $V_{\rm LSR}=-18\pm10\rm~km~s^{-1}$ of numerous filaments of either side of the expanding envelope of HB9, indicating a distance of $d=2\pm0.8\rm~kpc$. No spectra and images are presented in \citet{Lozinskaya81}, so it is difficult to justify where the velocities are measured. \citet{DOdorico77} presented measurements of the emission line ratios of HB9, but the spectral resolution is only $\Delta\lambda\approx8~\text{\AA}$, insufficient to accurately measure the radial velocity.

With our medium resolution MDM 1.3m/CCDS spectra, we measure the radial velocity at a few different locations of both HB9 (Table~\ref{table:HB9specfit}) and G159.2+3.3 (Table~\ref{table:G159specfit}). The velocity profiles of the two objects along the slits (with the slit location marked in Figs.~\ref{fig:HB9img} and \ref{fig:G159img}) are presented in Figs.~\ref{fig:HB9profiles}d and \ref{fig:G159profiles}d. The median radial velocity of the southern shell of HB9 is $V_{\rm LSR}\approx +1.9\rm~km~s^{-1}$, and spreads in a broad range of $V_{\rm LSR}\sim [-30 - +50]\rm~km~s^{-1}$ (see Tables~\ref{table:HB9specfit} and \ref{table:G159specfit} for the measured values of individual regions). We can see that the largest velocity dispersion appears at a distance of $\sim -1^\prime - -2^\prime$ (Fig.~\ref{fig:HB9profiles}d), which is more clearly shown in the 2D spectra appearing as the distorted line shape (Fig.~\ref{fig:HB92DSpec}). In this area, the slit happens to locate on another bright star and penetrates some diffuse filaments (Fig.~\ref{fig:HB9img}d). The large velocity dispersion of $\Delta V_{\rm LSR}\sim80\rm~km~s^{-1}$ may indicate the heating or broadening of the gas by the SNR shock. In this case, the large range of measured $V_{\rm LSR}$ is not inconsistent with any of the above radial velocity measurements.

The two observations of G159.2+3.3 show a clear systematic shift on the radial velocity of $\sim7.5\rm~km~s^{-1}$ (Fig.~\ref{fig:G159profiles}d). Since we reduce the data in an identical way and do not see a similar shift between the two observations of HB9 (Fig.~\ref{fig:HB9profiles}d), we think the shift in G159.2+3.3 is likely real, caused by the relative motion of the gas falling into the two neighbouring slits (Fig.~\ref{fig:G159img}d). Compared to HB9, the velocity dispersion of G159.2+3.3 is significantly smaller, typically in the range of $V_{\rm LSR}\sim [+2.5 - +20]\rm~km~s^{-1}$, with a median value of $\sim+11.7\rm~km~s^{-1}$. $V_{\rm LSR}$ of the two objects are thus overlapping with each other. We therefore cannot rule out the possibility that HB9 and G159.2+3.3 are located at the same distance solely based on the radial velocity criterion.

The FWHM of the lines in HB9 (median value $\sim51.5\rm~km~s^{-1}$; Fig.~\ref{fig:HB9profiles}e) is systematically higher than those in G159.2+3.3 (median value $\sim36.3\rm~km~s^{-1}$; Fig.~\ref{fig:G159profiles}e). The additional broadening of the lines could be explained as the shock heating or broadening. While the FWHM of the lines in G159.2+3.3 is comparable to or slightly higher than the radial velocity dispersion along the slits ($\Delta V_{\rm LSR}\lesssim20\rm~km~s^{-1}$), the FWHM of HB9 is typically significantly lower than the velocity dispersion ($\Delta V_{\rm LSR}\sim80\rm~km~s^{-1}$). This indicates that there is no significant global gas motion in the \ion{H}{II} region G159.2+3.3, while such a global motion could be significant in an SNR shocked region in HB9. We emphasise that our longslit is larger than the angular size of G159.2+3.3, so the measured velocity dispersion is largely representative of the whole \ion{H}{II} region. However, the slit just covers a small part of the SE shell of HB9, so the measured velocity dispersion is just a lower limit of the entire SNR, but should be representative of an SNR shocked region.

\begin{figure}[hbt]
\centering
\includegraphics[width=0.5\textwidth,trim={0in 0in 0.1in 0.3in},clip]{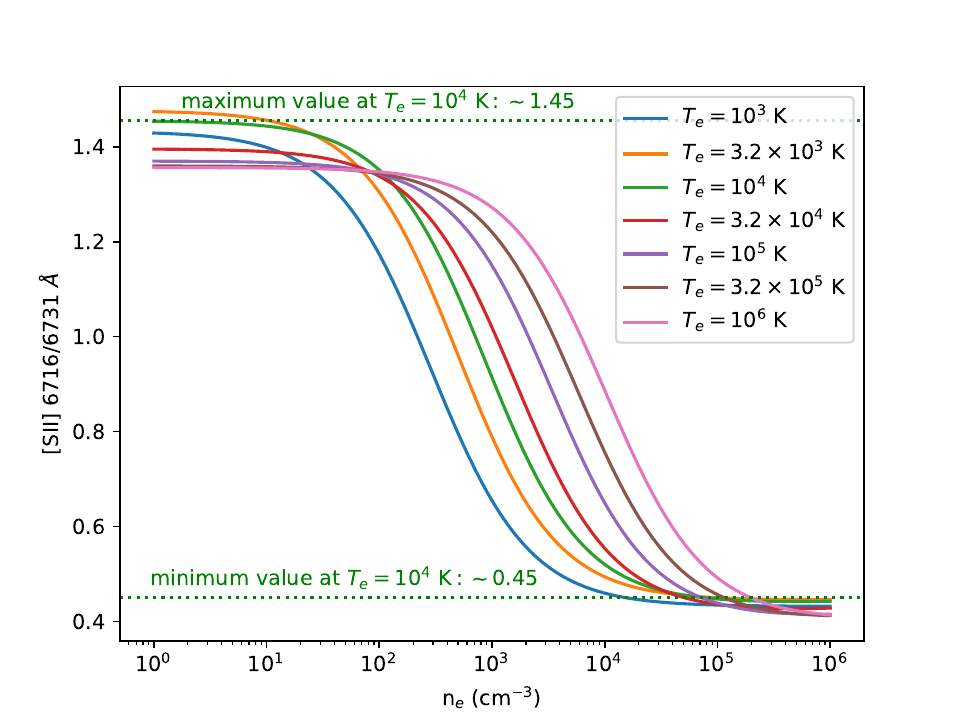}
\caption{Dependence of the [\ion{S}{II}] 6716/6730~\text{\AA} line ratio on the electron number density $n_{\rm e}$. Different curves are the model predictions at different temperatures computed using \texttt{PyNeb} \citep{Luridiana15}. The two dotted lines mark the approximate upper and lower physical limits on the [\ion{S}{II}] 6716/6730~\text{\AA} line ratio at $T_{\rm e}\sim10^4\rm~K$.}\label{fig:SIIRationeModel}
\end{figure}

\begin{figure}[hbt]
\centering
\includegraphics[width=0.5\textwidth,trim={0in 0in 0.1in 0.3in},clip]{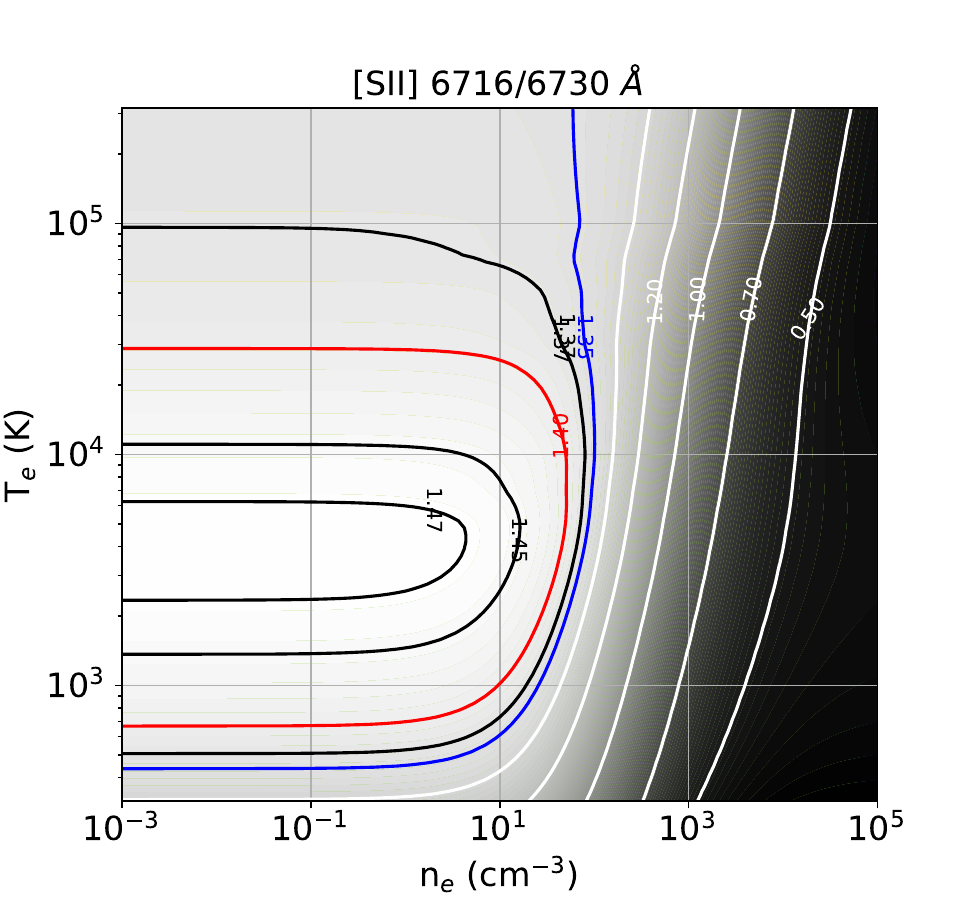}
\caption{Dependence of the [\ion{S}{II}] 6716/6730~\text{\AA} line ratio on the electron number density $n_{\rm e}$ and temperature $T_{\rm e}$. The grey scale image and contours are the model predictions computed using \texttt{PyNeb} \citep{Luridiana15}, with the corresponding line ratios denoted beside each contour. The blue (1.35) and red (1.40) contours are the median values measured in HB9 (Fig.~\ref{fig:HB9profiles}c) and G159.2+3.3 (Fig.~\ref{fig:G159profiles}c), respectively. For high density gas with $n_{\rm e}\gtrsim10^2\rm~cm^{-3}$, the [\ion{S}{II}] 6716/6730~\text{\AA} line ratio is highly density dependant so a good density tracer. On the other hand, at $n_{\rm e}\lesssim10^2\rm~cm^{-3}$, [\ion{S}{II}] 6716/6730~\text{\AA} is less sensitive to density, but more affected by temperature.}\label{fig:SIIRationeTe}
\end{figure}

\subsection{Line ratio and physical properties of ionised gas} \label{subsec:ionisedGas}

Physical parameters of the ionised gas could be measured with ratios of some emission lines (e.g., \citealt{Haffner99,Madsen06,Osterbrock06}). For example, the electron density $n_{\rm e}$ of the ionised gas could be measured with the ratio of two neighbouring density sensitive lines from the same ion, such as [\ion{S}{II}] 6716/6730, [\ion{O}{II}] 3729/3726, [\ion{Cl}{III}] 5517/5537, [\ion{Ar}{IV}] 4711/4740, \ion{C}{III}] 1906/1909, and [\ion{N}{1}] 5202/5199, which is not affected by the reddening, the ionisation state, or the metal abundance (e.g., \citealt{Copetti02}). The most commonly used ones are the strong [\ion{S}{II}] and [\ion{O}{II}] doublets, which are in general consistent with each other but still show some small systematic offsets (e.g., \citealt{Stanghellini89,Copetti02}). The [\ion{S}{II}] doublet is sensitive in a gas density range of $n_{\rm e}\sim10^{1-5}\rm~cm^{-3}$, corresponding to a [\ion{S}{II}] 6716/6730 line ratio of $\sim1.45-0.45$ (Fig.~\ref{fig:SIIRationeModel}). Using the [\ion{S}{II}] doublet, \citet{DOdorico77} obtained an average 6716/6730 line ratio of 1.33 and an electron density of $n_{\rm e}\approx690\rm~cm^{-3}$ toward the western shell of HB9 (Sect.~\ref{sec:Introduction}). 

Our medium-band CCDS spectra simultaneously cover a few emission lines including the [\ion{S}{II}] doublet (Sect.~\ref{subsec:CCDSData}; Figs.~\ref{fig:HB92DSpec}, \ref{fig:G1592DSpec}, \ref{fig:ExampleSpec}). After rejecting the spectra from a few regions with poor S/N and/or background subtraction so unphysical line ratios, we obtain a median [\ion{S}{II}] 6716/6730 line ratio of $\approx1.35$ for the SE shell of HB9 (Fig.~\ref{fig:HB9profiles}c; Table~\ref{table:HB9specfit}) and $\approx1.40$ for G159.2+3.3 (Fig.~\ref{fig:G159profiles}c; Table~\ref{table:G159specfit}). The latter is quite close to the upper limit expected from the $T_{\rm e}\sim10^4\rm~K$ gas (Fig.~\ref{fig:SIIRationeModel}). In order to estimate $n_{\rm e}$, we plot the [\ion{S}{II}] 6716/6730 line ratio as a function of both $n_{\rm e}$ and $T_{\rm e}$ in Fig.~\ref{fig:SIIRationeTe}, using the \texttt{PyNeb} code\footnote[2]{\url{http://research.iac.es/proyecto/PyNeb/}} \citep{Luridiana15}. For ionised gas at $T_{\rm e}>10^3\rm~K$, at a line ratio of $\mathcal{R}_{\rm 6716/6730}\approx1.35$ for the SE shell of HB9, the electron density is $n_{\rm e}\sim10^2\rm~cm^{-3}$ and insensitive to the temperature. This is roughly consistent with estimates from previous works in other parts of the SNR \citep{DOdorico77}. However, for G159.2+3.3, the [\ion{S}{II}] 6716/6730 line ratio of $\mathcal{R}_{\rm 6716/6730}\approx1.40$ is quite close to the physical upper limit, and depends on both $n_{\rm e}$ and $T_{\rm e}$. Assuming a typical ionised gas temperature of $T_{\rm e}\sim10^4\rm~K$, we can obtain $n_{\rm e}\sim50\rm~cm^{-3}$. At any temperatures, the derived $n_{\rm e}$ should be significantly lower than that in HB9.

The lower $n_{\rm e}$ of G159.2+3.3 than that of HB9, although estimated with large uncertainties, helps us to constrain the relative distance of the two objects. The flux density of an emission line (e.g., H$\alpha$; see the measured values in Tables~\ref{table:HB9specfit} and \ref{table:G159specfit}) in a thermally emitting nebula can be described as:
\begin{equation}
    F_{\rm H\alpha}\propto\frac{L}{d^2\theta^2}\propto\frac{n_{\rm e}^2Vf}{d^2\theta^2}\propto n_{\rm e}^2 d \theta f
\end{equation}
where $L$ is the luminosity of the object, $d$ is the distance, $\theta$ is the angular size, $V$ is the physical volume, and $f$ is the volume filling factor. As shown in both the images (Fig.~\ref{fig:HB9img}b) and the H$\alpha$ flux profiles (Figs.~\ref{fig:HB9profiles}a, \ref{fig:G159profiles}a), G159.2+3.3 has a significantly higher H$\alpha$ flux density than HB9, but $n_{\rm e}$ of G159.2+3.3 is indeed lower. For features with comparable angular sizes, G159.2+3.3 should have either a larger distance $d$, or a significantly larger filling factor $f$ than HB9. A larger distance to G159.2+3.3 (should be about one order of magnitude larger to compensate the lower $n_{\rm e}$, if all the measurements are accurate) seems to be consistent with its small angular size, although we cannot rule out the possibility that a small SF region locating at the same distance of the SNR ($\sim0.9-1.8\rm~pc$ assuming the distance of HB9 of $d\sim0.5-1\rm~kpc$; Sect.~\ref{subsec:VelocityDistance}). The same distance of G159.2+3.3 and HB9 is also not inconsistent with the radial velocity measurements (Sect.~\ref{subsec:VelocityDistance}) and the $\gamma$-ray morphology (Sect.~\ref{subsec:MultiwavImages}). On the other hand, if the SNR shock could significantly compress the ambient gas, as expected, we may expect a significantly higher $f$ in G159.2+3.3 than in the SE shell of HB9. This is an alternative explanation of the higher surface brightness but lower $n_{\rm e}$ of G159.2+3.3, but requires a fine tuning of $f$. Without any further evidence of the physical interaction between them, we prefer the two objects are not physically connected to each other.

The forbidden-to-Balmer line ratio is often a good indicator of the excitation and ionisation mechanisms. Based on our CCDS spectra, we also present the [\ion{N}{II}]/H$\alpha$ line ratio of the two objects in Figs.~\ref{fig:HB9profiles}b and \ref{fig:G159profiles}b. As the [\ion{N}{II}] 6548/6583 line ratio is almost a constant ($\approx1/3$), we herein only discuss the flux ratio between the stronger [\ion{N}{II}]~$\lambda6583~\text{\AA}$ line and the H$\alpha$ line. According to the ``Baldwin, Phillips \& Terlevich'' (BPT) diagram \citep{Baldwin81}, the median [\ion{N}{II}]~$\lambda6583~\text{\AA}$/H$\alpha$ line ratio of both the SE shell of HB9 ($\sim 0.48$) and G159.2+3.3 ($\sim 0.27$) could be classified as ``\ion{H}{II} regions'', indicating the ionisation sources are mainly hot stars, or in some cases also possibly to be low velocity shocks in SNRs (e.g., \citealt{Domcek2023}). However, at least in a few regions at a distance $\sim -0.5^\prime - 0^\prime$ from the slit acquisition star (${\rm distance}=0$) in HB9, we found some significantly elevated [\ion{N}{II}]~$\lambda6583~\text{\AA}$/H$\alpha$ line ratios up to $\gtrsim 1$, indicating additional ionisation sources. This sharp increase of the [\ion{N}{II}]~$\lambda6583~\text{\AA}$/H$\alpha$ line ratio is apparently associated with the post-shock area behind the bright SE shell (Fig.~\ref{fig:HB9img}d), so the elevated line ratio could be direct evidence of shock ionisation/excitation of the ambient gas, if the nitrogen abundance is not significantly elevated.

\section{Summary} \label{sec:Summary}

In this paper, we present optical observations of the Galactic SNR HB9 and the \ion{H}{II} region G159.2+3.3 projected close to it. The new H$\alpha$, H$\beta$, and [\ion{O}{III}] narrow-band images of G159.2+3.3 were taken with the MDM 1.3m telescope and the Templeton CCD. We also took medium-band medium-resolution spectra of a few selected regions in the SE shell of HB9 and G159.2+3.3 with the MDM 1.3m telescope and the CCDS spectrograph. Some prominent emission lines, such as  H$\alpha$~$\lambda6563~\text{\AA}$, [\ion{N}{II}]~$\lambda\lambda6548,6583~\text{\AA}$, [\ion{S}{II}]~$\lambda\lambda6716,6730~\text{\AA}$ are detected in many regions. We measure the flux, centroid velocity, and ratios of the lines, and study their spatial variation along the slits. 

The main results of this paper are summarised below:

1. We compared the optical narrow-band images to some multi-wavelength images, including our reproduced more than 15~yrs {\it Fermi}-LAT image at 1-300~GeV, the $^{12}$CO (J=1--0) image from the MWISP--CO line survey, the ROSAT X-ray images, the WISE IR images, the radio images at 1.4~GHz from NVSS, and at 4.85~GHz from GB6. For HB9, we found the H$\alpha$ filaments are in general spatially correlated with the radio image, indicating that they are both produced by the SNR shock. The X-ray emission, however, is centrally filled and enclosed by the H$\alpha$ and radio shells. This SNR is also bright in $\gamma$-rays, but the $\gamma$-ray morphology is not clearly associated with the surrounding molecular clouds, probably except for some molecular filaments in the velocity range of $[-5.5,-2.5]\rm~km~s^{-1}$ close to the SE shell. 

2. Based on our new narrow-band images of the \ion{H}{II} region G159.2+3.3, we found that the diffuse Balmer line emission has almost identical morphology as the radio emission, both enclosed by an IR bright shell, while the forbidden lines are quite weak. This multi-wavelength morphology is typical for Galactic \ion{H}{II} regions, which indicates that the Balmer line and radio emissions in the interior are thermal in origin, while the IR bright shell consists of dust heated by the UV photons from young stars.  

3. Based on our high-resolution spectra, we found that the radial velocity of the SE shell of HB9 sits in the range of $V_{\rm LSR}\sim[-30 - +50]\rm~km~s^{-1}$, consistent with previous measurements in other bands. This velocity dispersion over a small spatial scale indicates possible SNR shock heating of the gas. The velocity dispersion of G159.2+3.3 is significantly smaller, in the range of $V_{\rm LSR}\sim [+2.5 - +20]\rm~km~s^{-1}$. The FWHM of the lines in G159.2+3.3 is slightly higher than the radial velocity dispersion $\Delta V_{\rm LSR}$, while the FWHM of HB9 is significantly smaller than $\Delta V_{\rm LSR}$. This indicates additional global motion in HB9 caused by the SNR shock. 

4. We further calculated the electron density $n_{\rm e}$ from the [\ion{S}{II}] 6716/6730 line ratio. We found a median $n_{\rm e}\sim10^2\rm~cm^{-3}$ for the SE shell of HB9, while a smaller median $n_{\rm e}\sim50\rm~cm^{-3}$ for the brighter G159.2+3.3. This could be explained that the distance of G159.2+3.3 is indeed larger than that of HB9 so the two objects are not physically connected to each other, although we cannot rule out an alternative explanation which keeps the two objects physically connected but requires the filling factor of the shock compressed gas in the SNR significantly lower. 

5. The [\ion{N}{II}]~$\lambda6583~\text{\AA}$/H$\alpha$ line ratio of both G159.2+3.3 and most areas of the SE shell of HB9 covered by our spectral analysis can be interpreted with photo-ionisation by hot stars or low velocity shocks, such as in typical \ion{H}{II} regions. Only in the post-shock region in the SE shell of HB9, we found elevated [\ion{N}{II}]/H$\alpha$ line ratio which indicates possible contributions from shock ionisation.

In conclusion, a physical connection between the SNR HB9 and the compact \ion{H}{II} region G159.2+3.3 would require a significantly lower than unity filling factor of the ionised gas generated by shock compression in the SE shell of HB9. We cannot rule out this possibility, so the physical connection between the two objects remains plausible but unclear based on the current data.

\section*{Acknowledgements}

This research made use of \texttt{IRAF}, a general purpose software system for the reduction and analysis of astronomical data \citep{Tody86,Tody93}; \texttt{ccdproc}, an \texttt{Astropy} affiliated package for image reduction \citep{Craig17}; \texttt{PyNeb}, a lineage of tools dedicated to the analysis of emission lines \citep{Luridiana15}.
The authors acknowledge Eric Galayda at the MDM observatory for the help in setting up the MDM imaging and spectroscopy observations.
J.T.L acknowledges the financial support from the National Science Foundation of China (NSFC) through the grants 12273111 and 12321003, and also the science research grants from the China Manned Space Project.
Z.Q.X acknowledges the financial support from the Strategic Priority Research Programme of the Chinese Academy of Sciences No. XDB0550400 and the NSFC through the grant 12003069.
Y.C acknowledges the financial support from the NSFC through the grants 12173018, 12121003, \& 12393852.
P.Z acknowledges the financial support from the NSFC through the grant 12273010.
This research made use of the data from the Milky Way Imaging Scroll Painting (MWISP) project, which is a multi-line survey in 12CO/13CO/C18O along the northern galactic plane with PMO-13.7m telescope. We are grateful to all the members of the MWISP working group, particularly the staff members at PMO-13.7m telescope, for their long-term support. MWISP was sponsored by National Key R\&D Programme of China with grants 2023YFA1608000 \& 2017YFA0402701 and by CAS Key Research Programme of Frontier Sciences with grant QYZDJ-SSW-SLH047.

\bibliographystyle{aa}
\bibliography{G159+03}

\begin{thebibliography}{44}
\expandafter\ifx\csname natexlab\endcsname\relax\def\natexlab#1{#1}\fi

\bibitem[{{Abdollahi} {et~al.}(2020){Abdollahi}, {Acero}, {Ackermann}, {Ajello}, {Atwood}, {Axelsson}, {Baldini}, {Ballet}, {Barbiellini}, {Bastieri}, {Becerra Gonzalez}, {Bellazzini}, {Berretta}, {Bissaldi}, {Blandford}, {Bloom}, {Bonino}, {Bottacini}, {Brandt}, {Bregeon}, {Bruel}, {Buehler}, {Burnett}, {Buson}, {Cameron}, {Caputo}, {Caraveo}, {Casandjian}, {Castro}, {Cavazzuti}, {Charles}, {Chaty}, {Chen}, {Cheung}, {Chiaro}, {Ciprini}, {Cohen-Tanugi}, {Cominsky}, {Coronado-Bl{\'a}zquez}, {Costantin}, {Cuoco}, {Cutini}, {D'Ammando}, {DeKlotz}, {de la Torre Luque}, {de Palma}, {Desai}, {Digel}, {Di Lalla}, {Di Mauro}, {Di Venere}, {Dom{\'\i}nguez}, {Dumora}, {Fana Dirirsa}, {Fegan}, {Ferrara}, {Franckowiak}, {Fukazawa}, {Funk}, {Fusco}, {Gargano}, {Gasparrini}, {Giglietto}, {Giommi}, {Giordano}, {Giroletti}, {Glanzman}, {Green}, {Grenier}, {Griffin}, {Grondin}, {Grove}, {Guiriec}, {Harding}, {Hayashi}, {Hays}, {Hewitt}, {Horan}, {J{\'o}hannesson}, {Johnson}, {Kamae}, {Kerr}, {Kocevski}, {Kovac'evic'},
  {Kuss}, {Landriu}, {Larsson}, {Latronico}, {Lemoine-Goumard}, {Li}, {Liodakis}, {Longo}, {Loparco}, {Lott}, {Lovellette}, {Lubrano}, {Madejski}, {Maldera}, {Malyshev}, {Manfreda}, {Marchesini}, {Marcotulli}, {Mart{\'\i}-Devesa}, {Martin}, {Massaro}, {Mazziotta}, {McEnery}, {Mereu}, {Meyer}, {Michelson}, {Mirabal}, {Mizuno}, {Monzani}, {Morselli}, {Moskalenko}, {Negro}, {Nuss}, {Ojha}, {Omodei}, {Orienti}, {Orlando}, {Ormes}, {Palatiello}, {Paliya}, {Paneque}, {Pei}, {Pe{\~n}a-Herazo}, {Perkins}, {Persic}, {Pesce-Rollins}, {Petrosian}, {Petrov}, {Piron}, {Poon}, {Porter}, {Principe}, {Rain{\`o}}, {Rando}, {Razzano}, {Razzaque}, {Reimer}, {Reimer}, {Remy}, {Reposeur}, {Romani}, {Saz Parkinson}, {Schinzel}, {Serini}, {Sgr{\`o}}, {Siskind}, {Smith}, {Spandre}, {Spinelli}, {Strong}, {Suson}, {Tajima}, {Takahashi}, {Tak}, {Thayer}, {Thompson}, {Tibaldo}, {Torres}, {Torresi}, {Valverde}, {Van Klaveren}, {van Zyl}, {Wood}, {Yassine}, \& {Zaharijas}}]{Abdollahi20}
{Abdollahi}, S., {Acero}, F., {Ackermann}, M., {et~al.} 2020, \apjs, 247, 33

\bibitem[{{Ajello} {et~al.}(2021){Ajello}, {Atwood}, {Axelsson}, {Bagagli}, {Bagni}, {Baldini}, {Bastieri}, {Bellardi}, {Bellazzini}, {Bissaldi}, {Bloom}, {Bonino}, {Bregeon}, {Brez}, {Bruel}, {Buehler}, {Buson}, {Cameron}, {Caraveo}, {Cavazzuti}, {Ceccanti}, {Chen}, {Cheung}, {Ciprini}, {Cognard}, {Cohen-Tanugi}, {Cutini}, {D'Ammando}, {de la Torre Luque}, {de Palma}, {Digel}, {Dirirsa}, {Di Lalla}, {Di Venere}, {Dom{\'\i}nguez}, {Fabiani}, {Ferrara}, {Fiori}, {Foglia}, {Fukazawa}, {Fusco}, {Gargano}, {Gasparrini}, {Giroletti}, {Glanzman}, {Green}, {Griffin}, {Grondin}, {Grove}, {Guillemot}, {Guiriec}, {Gustafsson}, {Hays}, {Horan}, {J{\'o}hannesson}, {Johnson}, {Kamae}, {Kerr}, {Kuss}, {Larsson}, {Latronico}, {Lemoine-Goumard}, {Li}, {Liodakis}, {Longo}, {Loparco}, {Lovellette}, {Lubrano}, {Maldera}, {Manfreda}, {Mart{\'\i}-Devesa}, {Mazziotta}, {Menon}, {Mereu}, {Meyer}, {Michelson}, {Minuti}, {Mitthumsiri}, {Mizuno}, {Mongelli}, {Monzani}, {Moskalenko}, {Negro}, {Nuss}, {Ojha}, {Orienti}, {Orlando},
  {Paccagnella}, {Paliya}, {Paneque}, {Pei}, {Perkins}, {Pesce-Rollins}, {Pinchera}, {Piron}, {Poon}, {Porter}, {Primavera}, {Principe}, {Racusin}, {Rain{\`o}}, {Rando}, {Rani}, {Rapposelli}, {Razzano}, {Razzaque}, {Reimer}, {Reimer}, {Russell}, {Saggini}, {Saz Parkinson}, {Scolieri}, {Serini}, {Sgr{\`o}}, {Siskind}, {Smith}, {Spandre}, {Spinelli}, {Suson}, {Tajima}, {Thayer}, {Thompson}, {Tibaldo}, {Torres}, {Tosti}, {Valverde}, {Vigiani}, \& {Zaharijas}}]{Ajello21}
{Ajello}, M., {Atwood}, W.~B., {Axelsson}, M., {et~al.} 2021, \apjs, 256, 12

\bibitem[{{Anderson} {et~al.}(2014){Anderson}, {Bania}, {Balser}, {Cunningham}, {Wenger}, {Johnstone}, \& {Armentrout}}]{Anderson14}
{Anderson}, L.~D., {Bania}, T.~M., {Balser}, D.~S., {et~al.} 2014, \apjs, 212, 1

\bibitem[{{Anderson} {et~al.}(2011){Anderson}, {Bania}, {Balser}, \& {Rood}}]{Anderson11}
{Anderson}, L.~D., {Bania}, T.~M., {Balser}, D.~S., \& {Rood}, R.~T. 2011, \apjs, 194, 32

\bibitem[{{Araya}(2014)}]{Araya14}
{Araya}, M. 2014, \mnras, 444, 860

\bibitem[{{Baldwin} {et~al.}(1981){Baldwin}, {Phillips}, \& {Terlevich}}]{Baldwin81}
{Baldwin}, J.~A., {Phillips}, M.~M., \& {Terlevich}, R. 1981, \pasp, 93, 5

\bibitem[{{Chen} {et~al.}(2008){Chen}, {Seward}, {Sun}, \& {Li}}]{Chen08}
{Chen}, Y., {Seward}, F.~D., {Sun}, M., \& {Li}, J.-t. 2008, \apj, 676, 1040

\bibitem[{{Copetti} \& {Writzl}(2002)}]{Copetti02}
{Copetti}, M.~V.~F. \& {Writzl}, B.~C. 2002, \aap, 382, 282

\bibitem[{Craig {et~al.}(2017)Craig, Crawford, Seifert, Robitaille, Sip{\H o}cz, Walawender, Vin{\'{\i}}cius, Ninan, Droettboom, Youn, Tollerud, Bray, Walker, Janga, Stotts, G{\"u}nther, Rol, Bach, Bradley, Deil, Price-Whelan, Barbary, Horton, Schoenell, Heidt, Gasdia, Nelson, \& Streicher}]{Craig17}
Craig, M., Crawford, S., Seifert, M., {et~al.} 2017, astropy/ccdproc: v1.3.0.post1

\bibitem[{{Deharveng} {et~al.}(2005){Deharveng}, {Zavagno}, \& {Caplan}}]{Deharveng05}
{Deharveng}, L., {Zavagno}, A., \& {Caplan}, J. 2005, \aap, 433, 565

\bibitem[{{D'Odorico} \& {Sabbadin}(1977)}]{DOdorico77}
{D'Odorico}, S. \& {Sabbadin}, F. 1977, \aaps, 28, 439

\bibitem[{{Dom{\v{c}}ek} {et~al.}(2023){Dom{\v{c}}ek}, {Hern{\'a}ndez Santisteban}, {Chiotellis}, {Boumis}, {Vink}, {Akras}, {Souropanis}, {Zhou}, \& {de Burgos}}]{Domcek2023}
{Dom{\v{c}}ek}, V., {Hern{\'a}ndez Santisteban}, J.~V., {Chiotellis}, A., {et~al.} 2023, \mnras, 526, 1112

\bibitem[{{Fesen} {et~al.}(2024){Fesen}, {Drechsler}, {Strottner}, {Falls}, {Sainty}, {Martino}, {Galli}, {Ludgate}, {Blauensteiner}, {Reich}, {Walker}, {di Cicco}, {Mittelman}, {Morgan}, {Ettahar Kaeouach}, {Rupert}, \& {Benkhaldoun}}]{Fesen24}
{Fesen}, R.~A., {Drechsler}, M., {Strottner}, X., {et~al.} 2024, arXiv e-prints, arXiv:2403.00317

\bibitem[{{Gao} {et~al.}(2011){Gao}, {Han}, {Reich}, {Reich}, {Sun}, \& {Xiao}}]{Gao11}
{Gao}, X.~Y., {Han}, J.~L., {Reich}, W., {et~al.} 2011, \aap, 529, A159

\bibitem[{{Gerardy} \& {Fesen}(2007)}]{Gerardy07}
{Gerardy}, C.~L. \& {Fesen}, R.~A. 2007, \mnras, 376, 929

\bibitem[{{Green}(2019)}]{Green19}
{Green}, D.~A. 2019, Journal of Astrophysics and Astronomy, 40, 36

\bibitem[{{Green}(2022)}]{Green22}
{Green}, D.~A. 2022, Cavendish Laboratory, Cambridge, United Kingdom (available at "http://www.mrao.cam.ac.uk/surveys/snrs/")

\bibitem[{{Gregory} {et~al.}(1996){Gregory}, {Scott}, {Douglas}, \& {Condon}}]{Gregory96}
{Gregory}, P.~C., {Scott}, W.~K., {Douglas}, K., \& {Condon}, J.~J. 1996, \apjs, 103, 427

\bibitem[{{Haffner} {et~al.}(1999){Haffner}, {Reynolds}, \& {Tufte}}]{Haffner99}
{Haffner}, L.~M., {Reynolds}, R.~J., \& {Tufte}, S.~L. 1999, \apj, 523, 223

\bibitem[{{Jing} {et~al.}(2023){Jing}, {Han}, {Hong}, {Wang}, {Gao}, {Hou}, {Zhou}, {Xu}, \& {Yang}}]{Jing23}
{Jing}, W.~C., {Han}, J.~L., {Hong}, T., {et~al.} 2023, \mnras, 523, 4949

\bibitem[{{Leahy} \& {Aschenbach}(1995)}]{Leahy95}
{Leahy}, D.~A. \& {Aschenbach}, B. 1995, \aap, 293, 853

\bibitem[{{Leahy} \& {Tian}(2007)}]{Leahy07}
{Leahy}, D.~A. \& {Tian}, W.~W. 2007, \aap, 461, 1013

\bibitem[{{Lozinskaya}(1981)}]{Lozinskaya81}
{Lozinskaya}, T.~A. 1981, Soviet Astronomy Letters, 7, 17

\bibitem[{{Luridiana} {et~al.}(2015){Luridiana}, {Morisset}, \& {Shaw}}]{Luridiana15}
{Luridiana}, V., {Morisset}, C., \& {Shaw}, R.~A. 2015, \aap, 573, A42

\bibitem[{{Madsen} {et~al.}(2006){Madsen}, {Reynolds}, \& {Haffner}}]{Madsen06}
{Madsen}, G.~J., {Reynolds}, R.~J., \& {Haffner}, L.~M. 2006, \apj, 652, 401

\bibitem[{{Makai} {et~al.}(2017){Makai}, {Anderson}, {Mascoop}, \& {Johnstone}}]{Makai17}
{Makai}, Z., {Anderson}, L.~D., {Mascoop}, J.~L., \& {Johnstone}, B. 2017, \apj, 846, 64

\bibitem[{{McGlynn} {et~al.}(1998){McGlynn}, {Scollick}, \& {White}}]{McGlynn98}
{McGlynn}, T., {Scollick}, K., \& {White}, N. 1998, in New Horizons from Multi-Wavelength Sky Surveys, ed. B.~J. {McLean}, D.~A. {Golombek}, J.~J.~E. {Hayes}, \& H.~E. {Payne}, Vol. 179, 465

\bibitem[{{Newville} {et~al.}(2024){Newville}, {Otten}, {Nelson}, {Stensitzki}, {Ingargiola}, {Allan}, {Fox}, {Carter}, {Micha{\l}}, {Osborn}, {Pustakhod}, {Weigand}, {lneuhaus}, {Aristov}, {Glenn}, {Mark}, {mgunyho}, {Deil}, {Hansen}, {Pasquevich}, {Foks}, {Zobrist}, {Frost}, {Stuermer}, {Jaskula}, {Caldwell}, {Eendebak}, {Pompili}, {Hedegaard Nielsen}, \& {Persaud}}]{Newville24}
{Newville}, M., {Otten}, R., {Nelson}, A., {et~al.} 2024, {lmfit/lmfit-py: 1.3.1}

\bibitem[{{Oka} \& {Ishizaki}(2022)}]{Oka22}
{Oka}, T. \& {Ishizaki}, W. 2022, \pasj, 74, 625

\bibitem[{{Osterbrock} \& {Ferland}(2006)}]{Osterbrock06}
{Osterbrock}, D.~E. \& {Ferland}, G.~J. 2006, {Astrophysics of gaseous nebulae and active galactic nuclei}

\bibitem[{{Parker} {et~al.}(1979){Parker}, {Gull}, \& {Kirshner}}]{Parker79}
{Parker}, R.~A.~R., {Gull}, T.~R., \& {Kirshner}, R.~P. 1979, {An emission-line survey of the Milky Way}, Vol. 434

\bibitem[{{Saito} {et~al.}(2020){Saito}, {Yamauchi}, {Nobukawa}, {Bamba}, \& {Pannuti}}]{Saito20}
{Saito}, M., {Yamauchi}, S., {Nobukawa}, K.~K., {Bamba}, A., \& {Pannuti}, T.~G. 2020, \pasj, 72, 65

\bibitem[{{Sezer} {et~al.}(2019){Sezer}, {Ergin}, {Yamazaki}, {Sano}, \& {Fukui}}]{Sezer19}
{Sezer}, A., {Ergin}, T., {Yamazaki}, R., {Sano}, H., \& {Fukui}, Y. 2019, \mnras, 489, 4300

\bibitem[{{Stanghellini} \& {Kaler}(1989)}]{Stanghellini89}
{Stanghellini}, L. \& {Kaler}, J.~B. 1989, \apj, 343, 811

\bibitem[{{Terzian}(1965)}]{Terzian65}
{Terzian}, Y. 1965, \apj, 142, 135

\bibitem[{{Tody}(1986)}]{Tody86}
{Tody}, D. 1986, in Society of Photo-Optical Instrumentation Engineers (SPIE) Conference Series, Vol. 627, Instrumentation in astronomy VI, ed. D.~L. {Crawford}, 733

\bibitem[{{Tody}(1993)}]{Tody93}
{Tody}, D. 1993, in Astronomical Society of the Pacific Conference Series, Vol.~52, Astronomical Data Analysis Software and Systems II, ed. R.~J. {Hanisch}, R.~J.~V. {Brissenden}, \& J.~{Barnes}, 173

\bibitem[{{van den Bergh}(1978)}]{vandenBergh78}
{van den Bergh}, S. 1978, \apjs, 38, 119

\bibitem[{{van den Bergh} {et~al.}(1973){van den Bergh}, {Marscher}, \& {Terzian}}]{vandenBergh73}
{van den Bergh}, S., {Marscher}, A.~P., \& {Terzian}, Y. 1973, \apjs, 26, 19

\bibitem[{{Voges} {et~al.}(1999){Voges}, {Aschenbach}, {Boller}, {Br{\"a}uninger}, {Briel}, {Burkert}, {Dennerl}, {Englhauser}, {Gruber}, {Haberl}, {Hartner}, {Hasinger}, {K{\"u}rster}, {Pfeffermann}, {Pietsch}, {Predehl}, {Rosso}, {Schmitt}, {Tr{\"u}mper}, \& {Zimmermann}}]{Voges99}
{Voges}, W., {Aschenbach}, B., {Boller}, T., {et~al.} 1999, \aap, 349, 389

\bibitem[{{Yamauchi} \& {Koyama}(1993)}]{Yamauchi93}
{Yamauchi}, S. \& {Koyama}, K. 1993, \pasj, 45, 545

\bibitem[{{Zhao} {et~al.}(2020){Zhao}, {Jiang}, {Li}, {Chen}, {Yu}, \& {Wang}}]{Zhao20}
{Zhao}, H., {Jiang}, B., {Li}, J., {et~al.} 2020, \apj, 891, 137

\bibitem[{{Zhou} {et~al.}(2018){Zhou}, {Li}, {Zhang}, {Vink}, {Chen}, {Arias}, {Patnaude}, \& {Bregman}}]{Zhou18}
{Zhou}, P., {Li}, J.-T., {Zhang}, Z.-Y., {et~al.} 2018, \apj, 865, 6

\bibitem[{{Zhou} {et~al.}(2023){Zhou}, {Su}, {Yang}, {Chen}, {Sun}, {Jiang}, {Wang}, {Wang}, {Zhang}, {Xu}, {Yan}, {Yuan}, {Chen}, {Ao}, \& {Ma}}]{Zhou23}
{Zhou}, X., {Su}, Y., {Yang}, J., {et~al.} 2023, \apjs, 268, 61

\end{thebibliography}

\onecolumn 

\appendix

\section{Example spectra and spectral analysis results}\label{sec:expspec}

\begin{figure*}[hbt]
\centering
\includegraphics[width=1.0\textwidth]{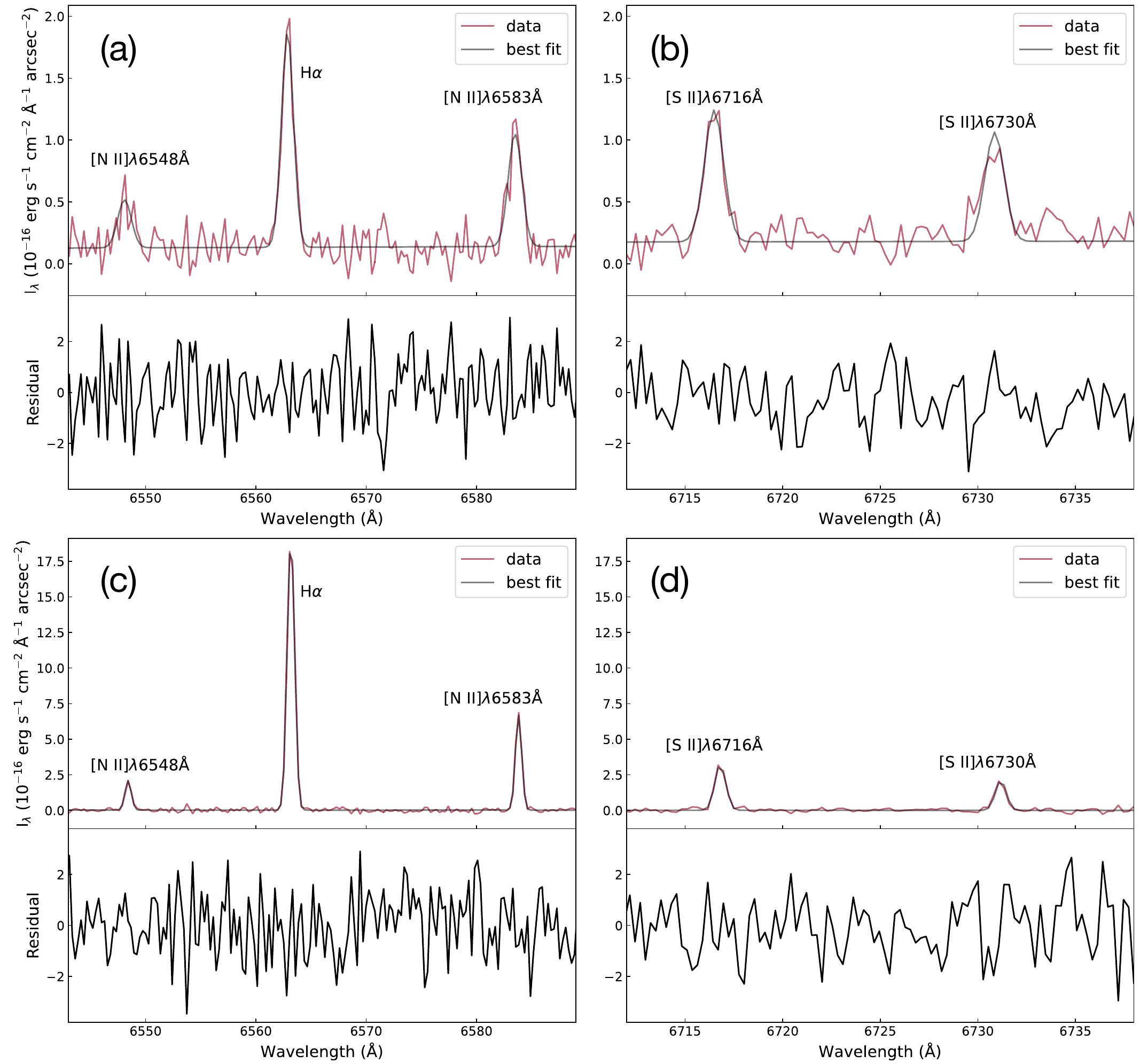}
\caption{Example spectra (red solid lines) of (a, b) HB9 extracted on 2024.03.29 at the position of the acquisition star ($d=0$, No.$=25$; Fig.~\ref{fig:HB9img}) and (c, d) G159.2+3.3 extracted on 2024.03.27 at $d=-2.25^{\prime}$ from the acquisition star (No.$=15$; Fig.~\ref{fig:G159img}). The grey solid curve is the best-fit to the spectra with Gaussian models plus a linear continuum. The lower panels show the spectral fit residuals defined as \it{(model - data)/error}. The slight difference in the residual continuum level in the upper and lower panels is caused by the imperfect local background modelling and subtraction, which however, does not affect the modelling of the emission lines. 
}\label{fig:ExampleSpec}
\end{figure*}

The appendix shows two example spectra extracted from the SE shell of HB9 (Fig.~\ref{fig:ExampleSpec}a) and the \ion{H}{II} region G159.2+3.3 (Fig.~\ref{fig:ExampleSpec}b), respectively. We also summarise spectral analysis results of different regions along the slits of the two objects in Tables~\ref{table:HB9specfit} and \ref{table:G159specfit}. In the tables, we only list results of the regions with significant emission line features detected.

\begin{table*}[ht!]
\caption{Spectral Analysis Results for HB9}\label{table:HB9specfit}
\centering
\footnotesize
\tabcolsep=2.0pt%
\begin{tabular}{lccccccccccc}
\toprule
\midrule
No. & $d$ & $v$ & FWHM & $F_{\rm H\alpha}$ & $F_{\rm [NII]6548}$ & $F_{\rm [NII]6583}$ & $F_{\rm [SII]6716}$ & $F_{\rm [SII]6730}$ & $\mathcal{R}_{\rm 6716/6730}$ & $n_{\rm e}$ \\
   & $^\prime$ & km/s & km/s & $10^{-16} \rm erg/s/cm^2/arcsec^2$ & $F_{\rm H\alpha}$ & $F_{\rm H\alpha}$ & $F_{\rm H\alpha}$ & $F_{\rm H\alpha}$ & & $\rm cm^{-3}$ \\
\midrule
Obs & Date & 2024.03.29 &  &  &  &  &  &  &  &  \\
\hline
        11 & -3.50 & $-11.62\pm 3.00$ & $28.28\pm 10.99$ & $0.50\pm 0.16$ & $<0.04$ & $0.44\pm0.27$ & $0.75\pm0.32$ & $0.58\pm0.35$ & $1.29\pm0.70$ & 167 \\
        12 & -3.25 & $-14.21\pm 2.30$ & $22.85\pm 8.23$ & $0.53\pm 0.17$ & $0.31\pm0.16$ & $0.57\pm0.29$ & $1.00\pm0.35$ & $0.66\pm0.31$ & - & - \\
        13 & -3.00 & $-7.02\pm 4.17$ & $25.81\pm 12.10$ & $0.60\pm 0.24$ & $0.34\pm0.19$ & $0.45\pm0.28$ & $<0.19$ & $<0.08$ & - & - \\
        15 & -2.50 & $-26.22\pm 3.36$ & $36.82\pm 9.17$ & $0.55\pm 0.14$ & $0.38\pm0.19$ & $0.49\pm0.22$ & $0.92\pm0.39$ & $0.54\pm0.23$ & - & - \\
        17 & -2.00 & $50.27\pm 3.58$ & $59.75\pm 12.40$ & $0.87\pm 0.16$ & $0.22\pm0.10$ & $0.50\pm0.15$ & $0.53\pm0.17$ & $0.38\pm0.15$ & $1.39\pm0.55$ & 55 \\
        18 & -1.75 & $25.48\pm 2.43$ & $58.52\pm 7.86$ & $1.40\pm 0.19$ & $0.15\pm0.11$ & $0.38\pm0.09$ & $0.26\pm0.07$ & $0.26\pm0.08$ & $1.02\pm0.36$ & 669 \\
        20 & -1.25 & $31.25\pm 8.48$ & $42.25\pm 36.04$ & $0.14\pm 0.11$ & $1.33\pm1.18$ & $<1.24$ & $<0.86$ & $<0.86$ & - & - \\
        22 & -0.75 & $-8.84\pm 3.69$ & $82.22\pm 9.54$ & $0.99\pm 0.11$ & $0.26\pm0.07$ & $0.20\pm0.07$ & $0.44\pm0.15$ & $0.29\pm0.22$ & - & - \\
        23 & -0.50 & $10.67\pm 1.96$ & $77.70\pm 7.73$ & $1.10\pm 0.13$ & $0.22\pm0.05$ & $0.90\pm0.13$ & $0.81\pm0.17$ & $0.75\pm0.15$ & $1.08\pm0.25$ & 514 \\
        24 & -0.25 & $7.80\pm 1.12$ & $41.88\pm 3.32$ & $1.40\pm 0.10$ & $0.47\pm0.06$ & $1.26\pm0.11$ & $1.22\pm0.19$ & $1.07\pm0.14$ & $1.15\pm0.20$ & 383 \\
        25 & 0.00 & $1.98\pm 1.23$ & $58.82\pm 4.26$ & $2.38\pm 0.19$ & $0.25\pm0.05$ & $0.60\pm0.08$ & $0.55\pm0.06$ & $0.46\pm0.07$ & $1.20\pm0.17$ & 288 \\
        26 & 0.25 & $2.06\pm 1.01$ & $69.96\pm 3.31$ & $4.36\pm 0.24$ & $0.10\pm0.02$ & $0.33\pm0.03$ & $0.33\pm0.03$ & $0.27\pm0.05$ & $1.21\pm0.23$ & 272 \\
        27 & 0.50 & $1.79\pm 0.80$ & $71.61\pm 1.85$ & $6.53\pm 0.22$ & $0.04\pm0.01$ & $0.11\pm0.01$ & $0.17\pm0.02$ & $0.12\pm0.03$ & $1.38\pm0.38$ & 73 \\
\hline
Obs & Date & 2024.03.30 &  &  &  &  &  &  &  &  \\
\hline
        11 & -3.50 & $-7.68\pm 2.21$ & $28.90\pm 8.68$ & $0.53\pm 0.14$ & $0.26\pm0.14$ & $0.31\pm0.17$ & $0.74\pm0.26$ & $0.36\pm0.22$ & - & - \\
        12 & -3.25 & $-8.16\pm 2.32$ & $32.30\pm 10.81$ & $0.50\pm 0.15$ & $<0.17$ & $0.86\pm0.34$ & $1.10\pm0.38$ & $0.77\pm0.33$ & $1.43\pm0.51$ & 22 \\
        13 & -3.00 & $2.19\pm 2.54$ & $36.99\pm 10.31$ & $0.63\pm 0.17$ & $0.22\pm0.14$ & $0.55\pm0.21$ & $0.47\pm0.20$ & $0.17\pm0.12$ & - & - \\
        14 & -2.75 & $-21.80\pm 4.13$ & $58.41\pm 13.49$ & $0.68\pm 0.16$ & $<0.16$ & $0.50\pm0.28$ & $0.74\pm0.26$ & $0.59\pm0.21$ & $1.27\pm0.47$ & 202 \\
        15 & -2.50 & $-15.71\pm 3.05$ & $51.65\pm 7.86$ & $1.11\pm 0.18$ & $<0.04$ & $0.60\pm0.16$ & $0.31\pm0.11$ & $0.45\pm0.12$ & $0.68\pm0.25$ & 2853 \\
        16 & -2.25 & $-26.03\pm 5.30$ & $39.15\pm 14.58$ & $0.33\pm 0.11$ & $0.59\pm0.42$ & $<0.40$ & $<0.19$ & $0.49\pm0.29$ & - & - \\
        17 & -2.00 & $50.74\pm 3.24$ & $48.88\pm 9.56$ & $0.83\pm 0.16$ & $<0.19$ & $0.78\pm0.21$ & $0.65\pm0.20$ & $0.31\pm0.16$ & - & - \\
        18 & -1.75 & $33.93\pm 2.00$ & $51.37\pm 5.52$ & $1.75\pm 0.19$ & $0.13\pm0.05$ & $0.48\pm0.09$ & $0.30\pm0.06$ & $0.21\pm0.07$ & $1.42\pm0.49$ & 29 \\
        19 & -1.50 & $47.77\pm 2.99$ & $42.81\pm 8.54$ & $1.01\pm 0.19$ & $0.11\pm0.05$ & $0.11\pm0.07$ & $0.17\pm0.07$ & $<0.13$ & - & - \\
        22 & -0.75 & $-0.16\pm 2.37$ & $81.51\pm 6.19$ & $1.49\pm 0.13$ & $0.28\pm0.05$ & $0.69\pm0.09$ & $0.38\pm0.09$ & $0.30\pm0.12$ & $1.27\pm0.54$ & 190 \\
        23 & -0.50 & $17.75\pm 0.86$ & $78.34\pm 3.49$ & $2.33\pm 0.09$ & $0.14\pm0.03$ & $0.54\pm0.04$ & $0.48\pm0.03$ & $0.22\pm0.03$ & - & - \\
        24 & -0.25 & $10.07\pm 1.49$ & $33.71\pm 4.67$ & $1.64\pm 0.23$ & $0.42\pm0.10$ & $1.42\pm0.24$ & $1.10\pm0.26$ & $0.99\pm0.21$ & $1.11\pm0.27$ & 443 \\
        25 & 0.00 & $9.67\pm 0.86$ & $59.55\pm 2.67$ & $3.57\pm 0.22$ & $0.14\pm0.02$ & $0.54\pm0.05$ & $0.50\pm0.05$ & $0.39\pm0.04$ & $1.29\pm0.15$ & 172 \\
        26 & 0.25 & $9.03\pm 0.81$ & $64.02\pm 3.03$ & $5.00\pm 0.30$ & $0.07\pm0.02$ & $0.36\pm0.03$ & $0.34\pm0.03$ & $0.25\pm0.04$ & $1.35\pm0.20$ & 104 \\
        27 & 0.50 & $6.10\pm 0.90$ & $72.56\pm 1.81$ & $9.06\pm 0.34$ & $0.03\pm0.01$ & $0.09\pm0.01$ & $0.04\pm0.01$ & $0.03\pm0.01$ & $1.22\pm0.48$ & 261 \\
\midrule
\bottomrule
\end{tabular}
\end{table*}

\begin{table*}[ht!]
\caption{Spectral Analysis Results for G159.2+3.3}\label{table:G159specfit}
\centering
\footnotesize
\tabcolsep=2.0pt%
\begin{tabular}{lccccccccccc}
\toprule
\midrule
No. & $d$ & $v$ & FWHM & $F_{\rm H\alpha}$ & $F_{\rm [NII]6548}$ & $F_{\rm [NII]6583}$ & $F_{\rm [SII]6716}$ & $F_{\rm [SII]6730}$ & $\mathcal{R}_{\rm 6716/6730}$ & $n_{\rm e}$ \\
   & $^\prime$ & km/s & km/s & $10^{-16} \rm erg/s/cm^2/arcsec^2$ & $F_{\rm H\alpha}$ & $F_{\rm H\alpha}$ & $F_{\rm H\alpha}$ & $F_{\rm H\alpha}$ & & $\rm cm^{-3}$ \\
\midrule
Obs & Date & 2024.03.27 &  &  &  &  &  &  &  &  \\
\hline
        10 & -3.50 & $15.51\pm 1.14$ & $37.67\pm 2.48$ & $3.65\pm 0.34$ & $<0.02$ & $0.26\pm0.04$ & $0.14\pm0.03$ & $0.07\pm0.07$ & - & - \\
        11 & -3.25 & $11.73\pm 1.68$ & $34.04\pm 2.93$ & $3.78\pm 0.40$ & $<0.01$ & $0.26\pm0.05$ & $0.15\pm0.04$ & $<0.04$ & - & - \\
        12 & -3.00 & $15.07\pm 0.75$ & $36.51\pm 1.50$ & $7.19\pm 0.60$ & $0.06\pm0.01$ & $0.25\pm0.04$ & $0.11\pm0.02$ & $0.07\pm0.02$ & - & - \\
        13 & -2.75 & $15.72\pm 0.64$ & $34.77\pm 1.53$ & $11.34\pm 0.71$ & $0.08\pm0.01$ & $0.29\pm0.03$ & $0.09\pm0.01$ & $0.05\pm0.01$ & - & - \\
        14 & -2.50 & $18.18\pm 0.55$ & $36.62\pm 1.40$ & $16.04\pm 1.23$ & $0.08\pm0.01$ & $0.32\pm0.03$ & $0.16\pm0.02$ & $0.09\pm0.01$ & - & - \\
        15 & -2.25 & $16.42\pm 0.53$ & $35.54\pm 1.34$ & $15.96\pm 1.15$ & $0.09\pm0.01$ & $0.30\pm0.03$ & $0.14\pm0.02$ & $0.09\pm0.01$ & - & - \\
        16 & -2.00 & $18.27\pm 0.59$ & $36.25\pm 1.67$ & $15.68\pm 1.31$ & $0.08\pm0.01$ & $0.29\pm0.03$ & $0.12\pm0.02$ & $0.09\pm0.01$ & $1.39\pm0.20$ & 58 \\
        17 & -1.75 & $18.21\pm 0.83$ & $33.43\pm 2.76$ & $11.82\pm 2.39$ & $0.08\pm0.02$ & $0.27\pm0.06$ & $0.11\pm0.03$ & $0.06\pm0.01$ & - & - \\
        18 & -1.50 & $14.37\pm 0.71$ & $34.34\pm 1.58$ & $11.24\pm 0.85$ & $0.07\pm0.01$ & $0.26\pm0.03$ & $0.08\pm0.02$ & $0.05\pm0.01$ & - & - \\
        19 & -1.25 & $13.55\pm 0.71$ & $33.89\pm 1.74$ & $12.13\pm 0.87$ & $0.08\pm0.01$ & $0.25\pm0.03$ & $0.08\pm0.02$ & $0.06\pm0.01$ & $1.34\pm0.34$ & 109 \\
        20 & -1.00 & $11.70\pm 0.59$ & $34.54\pm 1.40$ & $9.79\pm 0.60$ & $0.08\pm0.01$ & $0.30\pm0.03$ & $0.10\pm0.02$ & $0.07\pm0.01$ & - & - \\
        21 & -0.75 & $13.14\pm 0.44$ & $38.63\pm 0.99$ & $13.50\pm 0.78$ & $0.10\pm0.01$ & $0.30\pm0.03$ & $0.15\pm0.01$ & $0.10\pm0.01$ & - & - \\
        22 & -0.50 & $12.68\pm 0.39$ & $38.35\pm 0.91$ & $17.71\pm 0.84$ & $0.10\pm0.01$ & $0.31\pm0.02$ & $0.18\pm0.01$ & $0.13\pm0.02$ & $1.40\pm0.19$ & 46 \\
        23 & -0.25 & $11.53\pm 0.37$ & $39.15\pm 0.75$ & $19.17\pm 0.80$ & $0.09\pm0.01$ & $0.29\pm0.02$ & $0.17\pm0.01$ & $0.13\pm0.01$ & $1.36\pm0.17$ & 85 \\
        24 & 0.00 & $12.50\pm 0.35$ & $41.99\pm 0.60$ & $24.10\pm 0.90$ & $0.08\pm0.00$ & $0.26\pm0.02$ & $0.24\pm0.02$ & $0.19\pm0.02$ & $1.23\pm0.14$ & 244 \\
        25 & 0.25 & $14.98\pm 0.41$ & $37.13\pm 1.16$ & $37.76\pm 2.81$ & $0.08\pm0.01$ & $0.26\pm0.03$ & $0.16\pm0.02$ & $0.15\pm0.01$ & $1.06\pm0.10$ & 559 \\
        26 & 0.50 & $14.40\pm 0.46$ & $38.29\pm 1.04$ & $19.02\pm 1.48$ & $0.09\pm0.01$ & $0.30\pm0.03$ & $0.19\pm0.02$ & $0.14\pm0.03$ & $1.32\pm0.23$ & 135 \\
        27 & 0.75 & $14.29\pm 0.41$ & $39.22\pm 1.04$ & $14.32\pm 0.83$ & $0.10\pm0.01$ & $0.31\pm0.03$ & $0.20\pm0.02$ & $0.15\pm0.02$ & $1.38\pm0.17$ & 73 \\
        28 & 1.00 & $15.08\pm 0.42$ & $39.55\pm 1.03$ & $10.07\pm 0.52$ & $0.12\pm0.01$ & $0.34\pm0.03$ & $0.22\pm0.02$ & $0.17\pm0.02$ & $1.32\pm0.18$ & 138 \\
        29 & 1.25 & $15.30\pm 0.53$ & $38.10\pm 1.25$ & $7.73\pm 0.48$ & $0.12\pm0.01$ & $0.37\pm0.04$ & $0.23\pm0.04$ & $0.14\pm0.02$ & - & - \\
        30 & 1.50 & $17.27\pm 0.65$ & $33.60\pm 2.40$ & $4.74\pm 0.39$ & $0.16\pm0.02$ & $0.53\pm0.07$ & $0.38\pm0.06$ & $0.25\pm0.04$ & - & - \\
        31 & 1.75 & $18.78\pm 1.13$ & $38.98\pm 4.02$ & $1.66\pm 0.26$ & $0.20\pm0.05$ & $0.58\pm0.11$ & $0.20\pm0.06$ & $0.10\pm0.09$ & - & - \\
\hline
Obs & Date & 2024.03.30 &  &  &  &  &  &  &  &  \\
\hline
        11 & -3.50 & $6.89\pm 0.83$ & $30.70\pm 1.85$ & $7.96\pm 0.51$ & $0.02\pm0.02$ & $0.19\pm0.03$ & $<0.13$ & $<0.05$ & - & - \\
        12 & -3.25 & $9.04\pm 0.60$ & $34.81\pm 1.48$ & $12.79\pm 0.68$ & $0.05\pm0.01$ & $0.25\pm0.02$ & $0.12\pm0.02$ & $<0.02$ & - & - \\
        13 & -3.00 & $8.95\pm 0.54$ & $34.69\pm 1.10$ & $17.00\pm 0.68$ & $0.06\pm0.01$ & $0.26\pm0.03$ & $0.13\pm0.02$ & $0.04\pm0.02$ & - & - \\
        14 & -2.75 & $8.95\pm 0.54$ & $34.72\pm 1.10$ & $16.97\pm 0.69$ & $0.06\pm0.01$ & $0.26\pm0.03$ & $0.11\pm0.02$ & $0.04\pm0.02$ & - & - \\
        15 & -2.50 & $9.33\pm 0.37$ & $33.99\pm 0.86$ & $17.20\pm 0.58$ & $0.06\pm0.01$ & $0.25\pm0.02$ & $0.11\pm0.02$ & $0.07\pm0.01$ & - & - \\
        16 & -2.25 & $9.23\pm 0.49$ & $33.83\pm 1.04$ & $13.58\pm 0.57$ & $0.06\pm0.01$ & $0.28\pm0.03$ & $0.13\pm0.02$ & $0.09\pm0.01$ & $1.40\pm0.25$ & 48 \\
        17 & -2.00 & $6.20\pm 0.57$ & $31.96\pm 1.00$ & $11.95\pm 0.63$ & $0.06\pm0.01$ & $0.24\pm0.03$ & $0.05\pm0.02$ & $0.03\pm0.02$ & - & - \\
        18 & -1.75 & $4.71\pm 0.61$ & $30.13\pm 1.23$ & $11.92\pm 0.60$ & $0.08\pm0.02$ & $0.24\pm0.03$ & $0.03\pm0.01$ & $0.01\pm0.01$ & - & - \\
        19 & -1.50 & $3.82\pm 0.61$ & $30.75\pm 1.32$ & $10.29\pm 0.49$ & $0.08\pm0.02$ & $0.26\pm0.03$ & $0.04\pm0.01$ & $<0.01$ & - & - \\
        20 & -1.25 & $3.92\pm 0.48$ & $34.94\pm 1.09$ & $12.62\pm 0.46$ & $0.05\pm0.01$ & $0.27\pm0.02$ & $0.12\pm0.01$ & $0.11\pm0.02$ & $1.12\pm0.26$ & 422 \\
        21 & -1.00 & $4.00\pm 0.39$ & $37.88\pm 0.90$ & $18.46\pm 0.55$ & $0.10\pm0.01$ & $0.31\pm0.02$ & $0.10\pm0.01$ & $0.08\pm0.02$ & $1.29\pm0.28$ & 172 \\
        22 & -0.75 & $2.38\pm 0.37$ & $39.01\pm 0.84$ & $16.56\pm 0.58$ & $0.07\pm0.01$ & $0.26\pm0.01$ & $0.15\pm0.02$ & $0.12\pm0.02$ & $1.29\pm0.25$ & 165 \\
        23 & -0.50 & $4.48\pm 0.26$ & $40.88\pm 0.62$ & $24.75\pm 0.56$ & $0.06\pm0.01$ & $0.26\pm0.01$ & $0.18\pm0.01$ & $0.16\pm0.01$ & $1.14\pm0.11$ & 404 \\
        24 & -0.25 & $5.95\pm 0.24$ & $38.44\pm 0.58$ & $31.82\pm 0.68$ & $0.08\pm0.01$ & $0.26\pm0.01$ & $0.16\pm0.01$ & $0.13\pm0.01$ & $1.24\pm0.12$ & 239 \\
        25 & 0.00 & $6.30\pm 0.39$ & $36.86\pm 0.91$ & $20.75\pm 0.79$ & $0.08\pm0.01$ & $0.26\pm0.02$ & $0.14\pm0.01$ & $0.11\pm0.01$ & $1.22\pm0.15$ & 268 \\
        26 & 0.25 & $6.53\pm 0.47$ & $36.34\pm 1.12$ & $16.05\pm 0.80$ & $0.08\pm0.01$ & $0.27\pm0.02$ & $0.11\pm0.01$ & $0.11\pm0.02$ & $1.07\pm0.22$ & 526 \\
        27 & 0.50 & $7.61\pm 0.43$ & $37.73\pm 0.84$ & $9.31\pm 0.27$ & $0.08\pm0.01$ & $0.30\pm0.02$ & $0.17\pm0.04$ & $0.12\pm0.02$ & - & - \\
        28 & 0.75 & $5.94\pm 0.45$ & $35.89\pm 1.02$ & $7.48\pm 0.24$ & $0.09\pm0.02$ & $0.37\pm0.03$ & $0.20\pm0.04$ & $0.17\pm0.03$ & $1.17\pm0.29$ & 351 \\
        29 & 1.00 & $9.32\pm 0.88$ & $32.08\pm 2.42$ & $4.55\pm 0.41$ & $0.09\pm0.03$ & $0.43\pm0.07$ & $0.24\pm0.04$ & $0.27\pm0.07$ & $0.89\pm0.25$ & 1098 \\
        30 & 1.25 & $16.56\pm 1.82$ & $21.30\pm 6.01$ & $0.95\pm 0.27$ & $0.70\pm0.32$ & $1.09\pm0.38$ & $0.83\pm0.33$ & $0.74\pm0.48$ & $1.11\pm0.70$ & 451 \\
\midrule
\bottomrule
\end{tabular}
\end{table*}

\end{document}